# Evaporation limited spreading of ethanol on rectangular porous strips: an experimental and theoretical investigation

**Rampally Srirama Chandra Murthy**[a*] and **Navneet Kumar**[a]

[a] *Dept. of Mechanical Engineering, Indian Institute of Technology, Jammu, Jammu & Kashmir, 181221*

* *correspondence to* rampallysriram@gmail.com

**Abstract**

Wicking is a widely studied process in both natural and artificial systems. In many industrial applications, such as heat pipes, the wicking liquid evaporates to regulate temperature effectively. This study focuses on a simpler scenario where liquid ethanol climbs a vertically oriented filter paper (FP) under laboratory conditions, facilitating mass loss through evaporation and inducing cooling. Three filter papers with different permeability values were used, and three diagnostic methods – optical imaging, thermal imaging, and precision weighing – were employed to understand the dynamics of the process. The results showed a steady-state height $(L_c)$ significantly lower than Jurin's limit in all cases, indicating that evaporative mass loss, and not gravity, limits the process. For instance, the filter paper 1005FP, with a capillary radius of $59\mu m$ and an average pore size of $2.50\mu m$, would reach a Jurin's height of ~$9.6cm$ with ethanol if evaporation were not allowed. However, when evaporation occurred, the height reduced to ~$1.2cm$, an eightfold decrease; a similar reduction by a factor of $\sim 3$ was observed for 1004FP. Further, thermal imaging revealed a non-constant temperature distribution along the filter paper, with an unusual temperature inversion near the middle of the wicking liquid. This observation led to an improvement of the *Constant Evaporation Model* (CEM by Fries et al., 2008 [1]) by accounting for the nonlinear behavior of evaporation rates varying with vertical position. This new model termed the *Non-Constant Evaporation Model* (NCEM), tested two power-law relations for evaporation rates (referred to as $+\lambda$ and $-\lambda$ cases), both of which successfully captured the key features of the process. The predicted height-time (h-t) curves from NCEM matched the experimental data closely, with deviations in $(L_c)$ values within ±1%, compared to ~22% for CEM. Specifically, the deviation was ~0.8% for 1005FP and ~0.6% for 1001FP, showing a significant improvement over the CEM. However, neither model fully captured the process, suggesting that further refinements are needed. Recognizing that the evaporation rate is a key factor, two new non-dimensional numbers - Evaporation Height Number (EHN) and Evaporation Time Number (ETN) were proposed. The variation of EHN with ETN for four cases (three from this study and one from Fries et al., 2008 [1]) fell within a narrow range (0.80 to 1.00) in steady state, showing good consistency. In contrast, when using Fries et al.'s [1] dimensionless numbers, no clear trend emerged, reinforcing the need to include evaporation in the non-dimensionalization process. These new numbers (EHN and ETN) offer a step toward achieving universal scaling in evaporation-driven wicking processes. The study's findings contribute to a deeper understanding of wicking dynamics, with potential applications in thermal management systems such as wicks and heat pipes.

**Keywords**: Wicking, Porous Media, Evaporation, Volatile Liquids, Evaporation based scalings

## 1. Introduction:

Liquids climbing onto a porous surface either in vertical or in horizontal directions are commonly occurring events across various sectors. In such processes, the liquid's motion is driven by the inherent capillary pressure governed by the pore structure in the material and the entire dynamics is a result of the solid-liquid-gas interaction [2], [3], [4], [5] at the interface; leading eventually to the surface tension forces [6], [7]. In equilibrium, the balance of forces due to the capillary pressure and that of the surface tension on a spherical surface (with radius, r) yields the classical Young-Laplace (YL) [8] equation, $p_\alpha - p_\beta = \frac{2\sigma}{R_c}$, where $p_\alpha$ and $p_\beta$ are the pressure values on the internal and external curvatures of the spherical surface. This equation explains the maximum possible height reached by a liquid in a vertical capillary tube (termed as "Jurin's limit") from the liquid's free surface in a container. Also of interest is the rate of liquid rise dynamics in such a system. Lucas-Washburn [9] provided a theoretical framework for the rate of liquid spread in a horizontal capillary tube (of constant cross-sectional area), where they proposed a correlation between the wicking time (t) and the wicking height (h) as $h \propto t^{0.5}$. Note that in this configuration, the capillary forces drive the liquid while the viscous forces oppose this motion; gravitational forces being unimportant in this orientation. In a (horizontal or vertical) porous medium, the relation is seemingly different compared to that of the simplified case, i.e., a constant area of cross-section capillary tube and, thus, the Lucas-Washburn model was reported to be applicable only for short periods [10] in such cases. Further, this model failed to accommodate for the unsaturated or partially saturated situations as well as for the effect of different structural geometries [11]. Later, for a more complete understanding of liquid rise dynamics in a general system (say for a vertical case), the Lucas-Washburn equation was modified to account for the inertia and gravity terms [12]. More recently, investigations on the influence of various forces on liquid rise dynamics were reported [13], [14], [15] to have three distinct regimes. A short inertial regime in the early stage (where the liquid forms the meniscus) followed by the 2nd stage where the viscous forces compete with the capillary forces (the original Lucas-Washburn scenario) and eventually in the 3rd stage, the gravitational forces become important, and limits the height rise known as Jurin's limit or YL height. Some of these investigations used water, milk, etc. in their experiments [16], [17] of liquid wicking on thin porous strips like a filter paper. Incidentally, the same law holds for a liquid rising in a corner [18].

The overall dynamics of liquid rise, especially in a porous network, becomes more complicated in the scenarios where phase change of the liquid into the liquid vapours [19], [20], [21] occurs (or desired such as in heat pipes) like water seepage in the walls of a building where the source maintains water seepage. The rate of evaporation becomes an important parameter in these processes. With liquid water as the working liquid, the investigation by Zhi et al. [22] reported a small deviation from the case with no evaporation. With considerably volatile liquids, however, the dynamics changes significantly, and one such study was undertaken by Fries and his research team [1]. The liquid used in one of these studies [1] was that used quite frequently in the printing industries, while the porous medium chosen was a metallic weave. They extended the Lucas-Washburn law by including an evaporation rate related term that eventually gets reflected as a part of the total viscous resistance term through what they term as the '*refill velocity*' (more discussion on this in Section 5). The most significant outcome from their research was the limited capillary rise of the chosen liquid on the metallic weave and that primarily was determined by the evaporative loss term. Though qualitatively similar, their theoretical model overestimated the experimental results by ~35%. This deviation may have been due to several factors like (a) the assumption of a constant evaporation rate throughout the height of the wicked liquid (from the bottom end), (b) the simplified Lucas-Washburn model (where capillary radius was experimentally obtained), (c) medium homogeneity, (d) one-dimensional treatment, etc. [1] . The rate of evaporation was estimated by the weighing balance data in Fries et al., 2008 [1]. In another recent investigation [23], the investigators avoided this mass measurement method and relied on the diffusion theory to calculate the transient evaporation rate at all the time instants that was incorporated as the latent heat term in the energy equation which was solved simultaneously with the mass and momentum equations. However, they seem to have misrepresented the temperature and mass coupling.

In this study, we use three simultaneous diagnostics – mass loss measurement, surface temperature measurements, and optical imaging – to understand this dynamic process in more detail. The surface temperature data was used to construct the surface energy budget (SEB) in the steady state to obtain the rate of evaporation as a function of the known / measured parameters such as surface temperature, ambient temperature, etc. We use the energy equation into the momentum equation, the solution of which gives t(h) if capillary radius and permeability is known. The variable power law-like relation between the rate of evaporation along the wicked liquid length (called NCEM hereafter) was used to capture the non-linear trend of the obtained surface temperature. We show that these key additions to the existing theoretical framework [1] captures the wicking dynamics involving the phase change in a better way. Extensive set of experiments were conducted to test and validate the proposed NCEM framework.

The article is arranged as follows. We first discuss the materials used in this study followed by the experimental methodology adopted in Section 2. Next, we show that the existing theory [1] is insufficient to capture this complex phenomenon (incorporating evaporation in the process of a liquid rising on a porous medium). In the subsequent section, we propose a better mathematical model to capture some more aspects; in particular, the height-wise variation of the rate of evaporation from the filter paper. The results and discussions are detailed in Sections 3,4 and 5 followed by conclusions in Section 6.

**Nomenclature**

| | | | |
|---|---|---|---|
| Symbols | | | |
| $a_0, a_1$ | Evaporation rate constant (kgm$^{-2}$s$^{-1}$) | $T_\infty$ | Ambient temperature (K) |
| $FP$ | Filter paper | $T_s$ | Surface temperature (K) |
| $A_b$ | Cross-sectional area of the FP (m$^2$) | $U$ | Uncertainty |
| $g$ | Gravitational acceleration (ms$^{-2}$) | $t$ | Time (s) |
| $h$ | Capillary penetration length (m) | $t_p$ | Thickness of the FP (m) |
| $\bar{h}$ | Average convective heat transfer coefficient (Wm$^{-2}$K$^{-1}$) | $v$ | Liquid front velocity (ms$^{-1}$) |
| $h_{fg}$ | Latent heat of vaporization (Jkg$^{-1}$) | $v_r$ | Refill velocity (ms$^{-1}$) |
| $H_{YL}$ | Jurin's limit (m) | $W$ | Width of the filter paper (m) |
| $K$ | Permeability (m$^2$) | | |
| $k$ | Thermal conductivity (Wm$^{-1}$K$^{-1}$) | Subscripts | |
| $Lc$ | Steady state length (m) | $opt$ | Optical resolution |
| $\dot{M}_e$ | Total evaporation mass flow (kgs$^{-1}$) | $res$ | Thermal resolution |
| $\dot{m}_e$ | Evaporated mass (kgm$^{-2}$s$^{-1}$) | $sen$ | sensitivity |
| $\dot{m}_g$ | Mass gain rate (kgs$^{-1}$) | $d$ | Dipped length (m) |
| $\dot{M}_h$ | Mass flow of the liquid front (kgs$^{-1}$) | Greek letters | |
| $\dot{m}_l$ | Mass loss rate (kgs$^{-1}$) | $\lambda$ | Fitting parameter |
| $\overline{Nu}$ | Average Nusselt number | $\mu$ | Dynamic viscosity (kgm$^{-1}$s$^{-1}$) |
| $p_c$ | Capillary pressure (Nm$^{-2}$) | $\phi$ | Porosity |
| $p_h$ | Hydrostatic pressure (Nm$^{-2}$) | | |
| $\dot{p}_h$ | Viscous pressure loss due to liquid front velocity ($Nm^{-2}$) | $\rho_l$ | Liquid density ($kgm^{-3}$) |
| $Pr$ | Prandtl number | | |
| $p_r$ | Viscous pressure loss due to refill velocity ($Nm^{-2}$) | $\sigma$ | Interfacial Surface tension ($Nm^{-1}$) |
| $Ra$ | Rayleigh number | $\sigma_s$ | Stefan-Boltzmann constant ($Wm^{-2}K^{-4}$) |
| $R_c$ | Capillary radius (m) | $\theta$ | Contact angle ( ° ) |
| | | $\epsilon$ | Emissivity |

## 2. Methods & Methodology

We used three types of thin filter papers (FP) and ethanol as the working liquid for the experimental study. The details are discussed next.

### 2.1 Porous medium and liquid properties:

We used Whatman Grade 1001, 1004, and 1005 to study liquid rise on `vertical thin rectangular paper strips, essentially a 2D porous medium. The geometrical dimensions and porosity values are seen in Table 1[24].

Table 1 Filter paper properties obtained from Whatman filter paper guide [24].

| Grade | Capillary Radius $R_c$ (m) | Paper Thickness $t_p$ (m) | Porosity $\phi$ |
|---|---|---|---|
| 1001 | 64 x $10^{-6}$ | 180 x $10^{-6}$ | 0.48 |
| 1004 | 75 x $10^{-6}$ | 205 x $10^{-6}$ | 0.76 |
| 1005 | 59 x $10^{-6}$ | 200 x $10^{-6}$ | 0.54 |

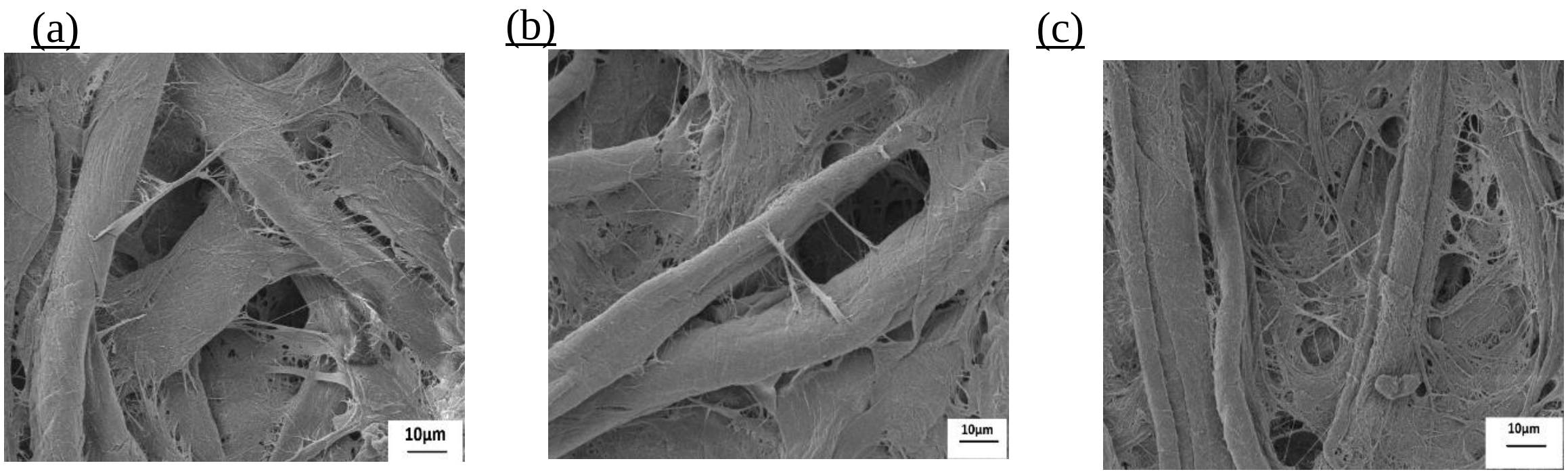

Figure 1 SEM images of Whatman grade (a) 1001FP, (b) 1004FP, and (c) 1005FP.

All the filter papers were cut precisely in a rectangular shape of dimension 8.5cm × 2.5cm using a laser cutter. Figure 1 shows images of these papers captured under a scanning electron microscope (SEM); it is evident that 1004FP is more porous, i.e., highest average pore radius compared to the other two FP's. The majority of experiments were performed with ethanol. Some experiments are performed with DI water, which can be treated as a non-volatile liquid up to certain extent. The idea is to understand the competition between surface tension driven capillary rise (working in favour of mass gain), gravitational forces (working against mass gain), viscous resistance (working against mass gain), and evaporation (considered as mass loss). Essentially, it can also be considered as the competition between the respective governing velocity scales (more clearly shown in Figure 12 and discussed therein).

## 2.2 Experimental Setup

Figure 2a and 2b show the schematic and a sample photograph of the experimental setup, respectively. A vertically suspended FP was centrally positioned above a container containing ethanol. The FP outside the liquid reservoir is exposed to the atmosphere facilitating evaporation. The bottom (~5mm) end of the FP was dipped into the pool of ethanol liquid while the top end was glued to another surface to maintain the vertical orientation of the FP (see Figure 2b). The top of the container was sealed, except for a small opening/slot at its top (Figure 2c) that allowed the FP to cross through such that the lower part of the FP could be submerged in the liquid. This configuration on one hand allows for seamless liquid ethanol rise on the FP and on the other hand restricts unwanted mass loss (via evaporation) from the liquid reservoir. Out of this 0.9cm$^2$ of open area on the top cover (see Figure 2c), ~0.5cm$^2$ is occupied by the cross-section of the FP. Hence, the area available for unwanted evaporation is negligible when compared to the wet liquid area on the FP (to be discussed soon). Utmost care was taken during the experiment so that the FP did not touch any wall of the slot or top cover. The ethanol container was connected to a source tank via a siphon tube. Both the source and container are made up of polymethyl methacrylate (PMMA) and are connected with a Polyurethane (PU) tube. A breather hole was provided on the top of surface tank (sealed from all the sides) in order to maintain the liquid surface pressure atmospheric.

The cross-sectional area of the container ($A_s$) and the source tank ($A_r$) are 100cm$^2$, and 25cm$^2$, respectively. This provision replenishes the liquid ethanol (from the source tank to the container) while evaporative mass loss continues to occur in the container; this allows for a longer duration of measurement while maintaining the liquid level in the container nearly the same. For example, if the source tank is not connected to the reservoir and the mass lost is 5gm, the source tank height would decrease by ~2mm for the case of water ($\Delta h_{level} = \frac{\Delta m}{\rho_w * A_{c/s}}$ $i.e., \Delta h_{source} = 2mm$). If a reservoir is connected to the source tank, the height reduction in the source tank is only 0.4 mm. The container and the source tank were placed on a platform on top of the weighing scale (discussed next).

$$\frac{h_{container}}{h_{container+source}} = 1 + \frac{A_s}{A_{s+r}} = 5, h_{with\ siphon\ setup} = \frac{2}{5} = 0.4mm\ .$$

Note that without replenishment, ~2mm reduction in water level in the container would correspond to ~24 hours of evaporation (assuming ~2mm/day [25] value in the ambient conditions without external heating and wind). With replenishment, this measurement time would be even longer without a significant drop in the liquid level. The experiments were conducted in a controlled laboratory conditions (temperature $27 \pm 0.5$ °C and relative humidity of $52 \pm 2\%$). Coming to the measurement part, we use three simultaneous diagnostic tools to understand the entire process; these are an optical camera, a thermal camera and a precision weighing scale (see Figure 2a,b). An optical camera (Nikkor Z5, 24MP) provides high-resolution visualization of the rising wicking front. A thermal camera (Fluke TiX 580, resolution of 640 × 480 pixels and a sensitivity of 0.05K) was used to measure the surface temperature; this camera can differentiate between two pixels that are 0.05K different in temperature. A precision weighing balance (SARTORIUS, model QUINTIX6102-10IN, least count of 0.01g) was used to measure the instantaneous mass loss from the system under consideration. Data from the weighing scale was logged into a computer at an interval of 10 seconds. Further, we use Surface Energy Budget (SEB) method to link the evaporative mass loss with the measured parameters – surface temperature of the FP, ambient temperature, wicking length, and convective heat transfer coefficient. The experiments were repeated three times to obtain the consistency across the experiments while a separate uncertainty analysis (more details in Appendix B) yields the error band in each experiment.

Figure 2d is a cartoon that shows the experimental configuration highlighting the important geometrical features. The total length of the FP is 8.5cm, out of which ~0.5cm is dipped in the liquid container, ~0.2cm is the top cover plate, and ~7.8cm length is exposed to the atmosphere and is outside the container; the optical camera is set up in such a way that it sees only this exposed part of the FP. Along this FP's exposed length, the number of pixels seen range from 840 to 870; FP's width is not important for the current investigation. The images are analyzed using the commercially available *ImageJ* software for precisely tracking the motion of the liquid ethanol front (the image processing is discussed in Section 3).

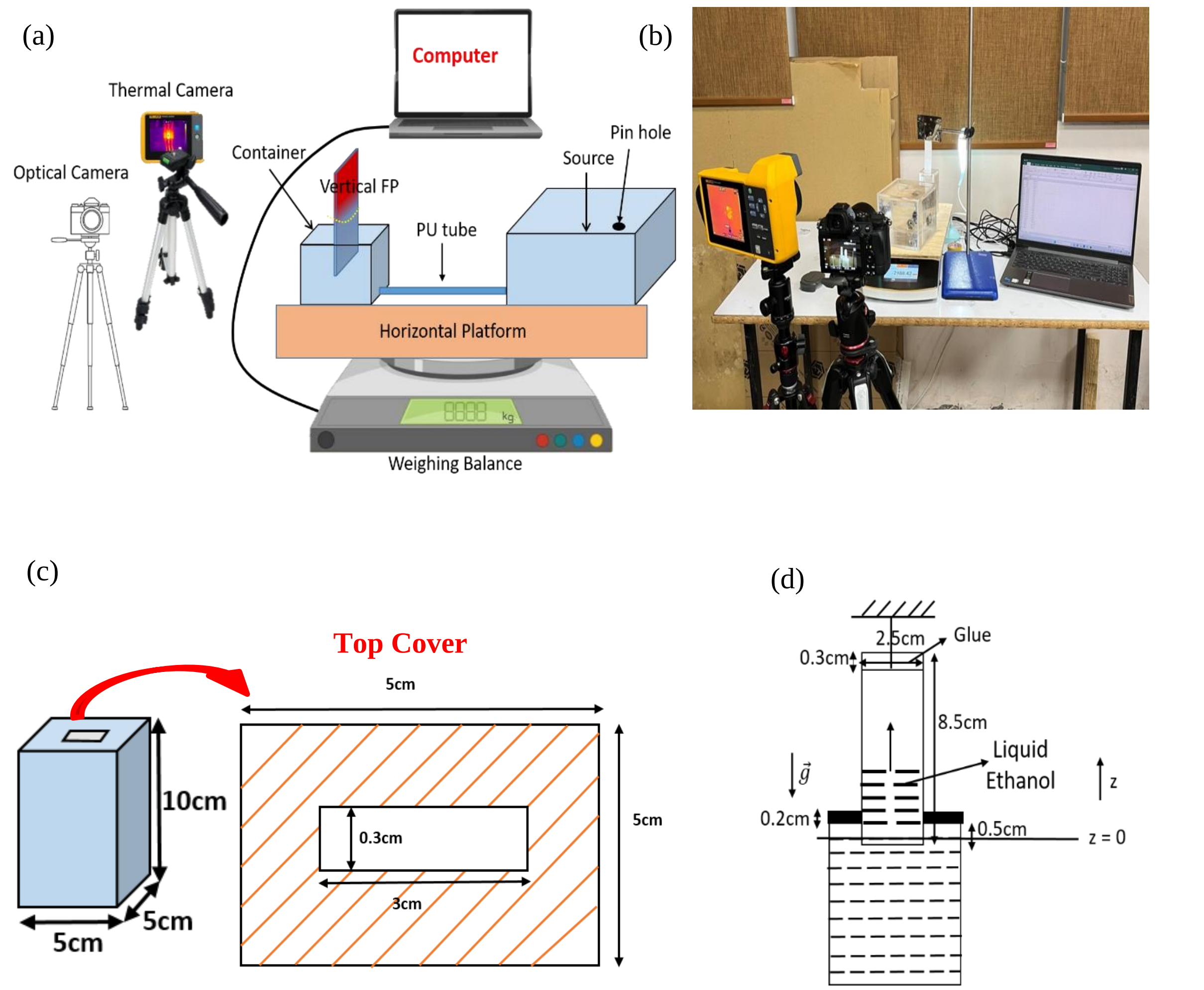

Figure 2 [Colour online] (a) Schematic of the experimental setup, (b) a sample photograph showing the experimental setup, (c) cartoons showing the design and geometry of the source tank and the top cover with a slot (3cm × 0.3cm) at its centre through which the FP passes, and (d) a cartoon illustrating the experimental setup, showing the FP's important geometrical features and the portion visible to the optical camera. The images, analyzed with ImageJ software, track the motion of the liquid ethanol front; *z* is positive in the upward direction and z = 0 marks the location where the FP meets the liquid pool kept in the container.

## 3. Results and discussion

### 3.1 Optical analysis

We first show results of variation of the liquid front height penetrated into the FP with time obtained through the optical camera and processed subsequently using image processing. Then, we discuss the temperature evolution in all the FP cases in the transient state as well as in the steady state. Here, we also discuss nature of different temperature profiles that form the basis of our mathematical modelling.

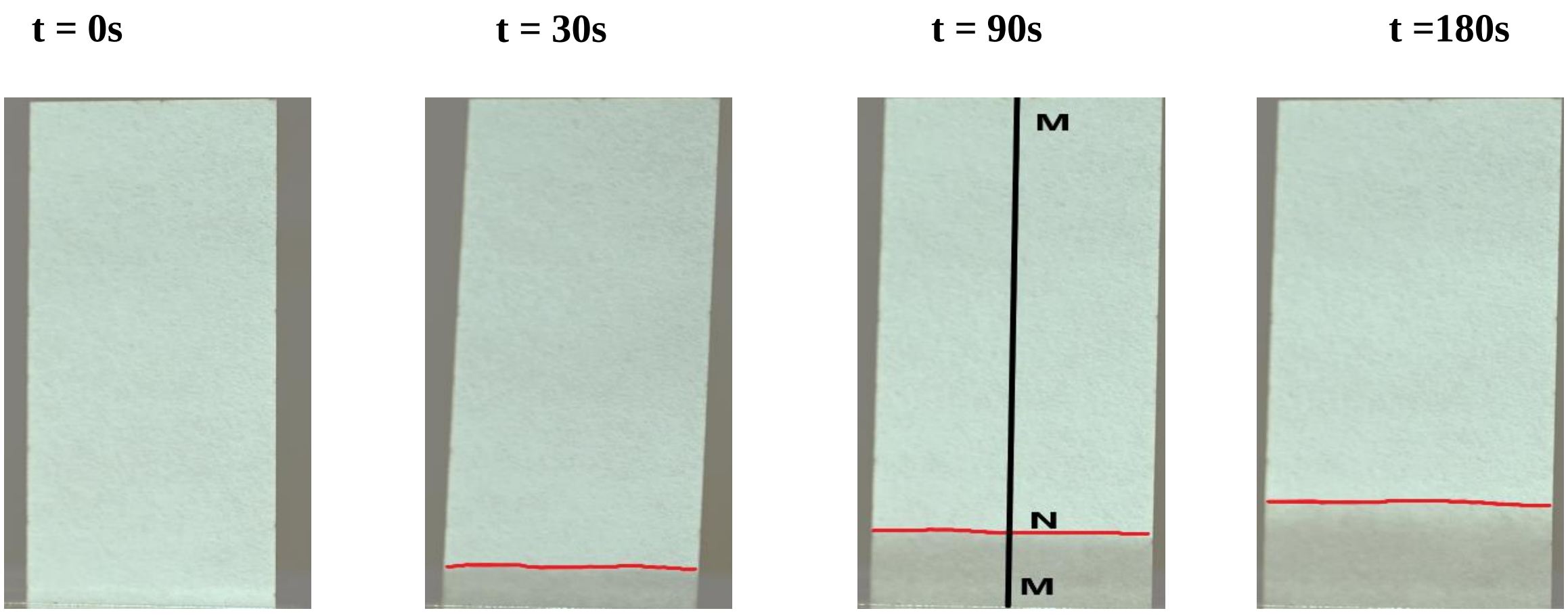

Figure 3 A series of snapshots captured at different time instants showing the motion of the liquid ethanol front on the 1004FP.

Figure 3 shows a series of snapshots (taken out of the recorded film) for the case of Whatman 1004FP at notable instances. Time t = 0 second is chosen when the liquid-vapor contact line in the FP begins to be seen in the optical camera; this is at the same level as the top of the 'Top cover'. Note that below this level, the development is not visible. Time t = 180 seconds corresponds to the stage very near to the steady state position of the liquid front. At t ~ 300 seconds (not shown in Figure 3), the process of liquid rise is complete (see Figure 4) and the liquid front stops growing. The images were subsequently analyzed using the commercially available *ImageJ* software to get the L-V meniscus locations at different times. The L-V meniscus position is tracked on the line MM (small inclination does not alter the measurements since this line itself is parallel to the papers' vertical edges) and the length MN was obtained; this gives the distance travelled by the liquid front (vertical distance calibration was done using a graph paper kept adjacent to the FP, in some cases and by measured FP length exposed to the ambient in other cases). Note that the camera window along the FP length is ~870 pixels and that corresponds to ~7.8cm length. One pixel, hence, leads to an error of ~0.10mm and this is the uncertainty $(U_{opt})$ in the adopted image processing. The collection of such extensive image analysis yields liquid rise height versus time as seen in Figure 4.

Figure 4 shows the temporal evolution of capillary penetration length $(h)$ for ethanol in all three FP cases studied. In all the cases, a rapid increase in *h* is seen initially (~20 seconds) and then the *h*-*t* curves follow a non-linear trend eventually reaching the corresponding steady state values (denoted by $L_c$); these are ~26mm, ~15mm, and ~12mm for 1004FP, 1001FP, and 1005FP cases, respectively. The maximum $L_c$ value is observed in the case of 1004FP, while the lowest is seen in the case of 1005 grade FP. The bars seen in Figure 4 represent the deviation across three experiments in each case.

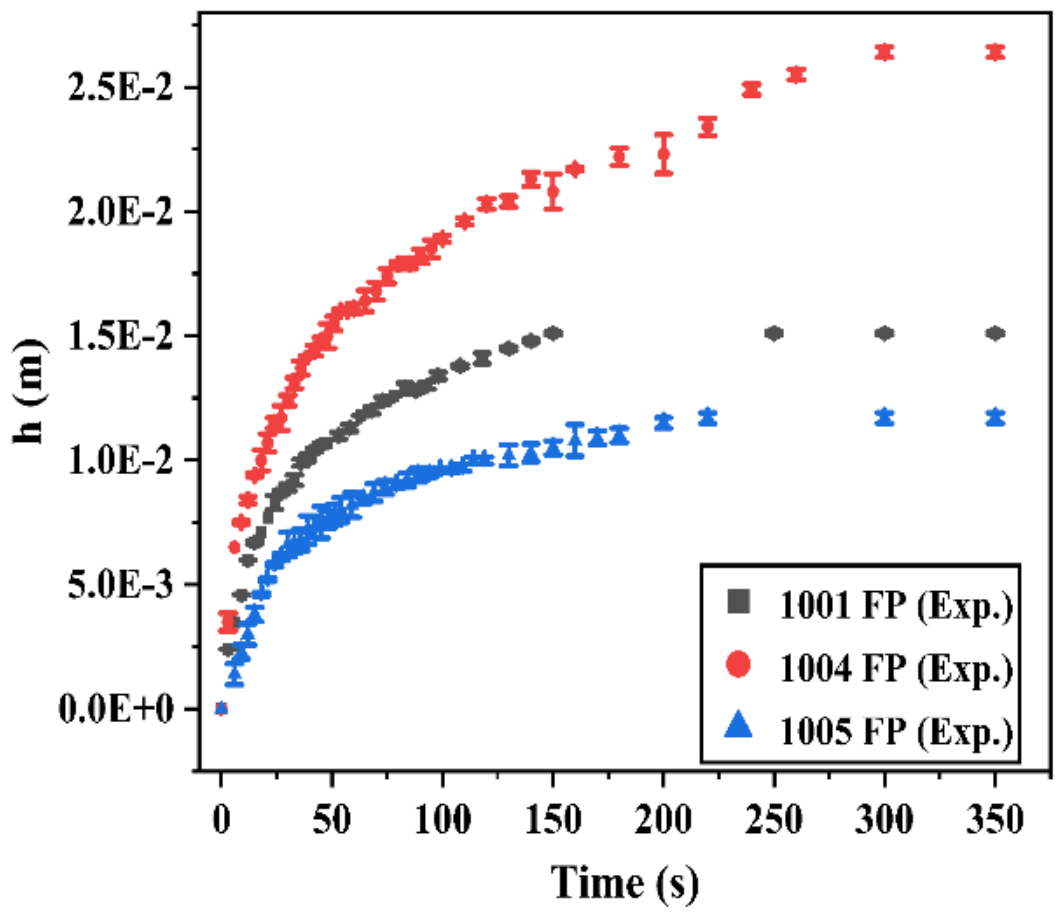

Figure 4 [Colour online] Temporal variation of capillary penetration length for liquid ethanol in three different filter papers. The bar represents the experimental deviation across three experiments in each case.

As seen in Table 2, if evaporation does not happen from the system, the L-V meniscus would reach the respective $H_{YL}$ values (Jurin's limit), which are much higher than the corresponding $L_c$ values. For the same liquid, $H_{YL}$ increases when $R_c$ decreases but $L_c$ decreases when $R_c$ is decreased (and this seems to be connected to 'Permeability'; to be discussed soon). Also, $H_{YL}/R_c$ values increase much faster compared to the $L_c/R_c$ values as '$R_c$' decreases. It seems clear that in the current investigation, gravitational force, is not the 'limiting' parameter (unlike Jurin's limit) at least when the liquid used is volatile in nature, thereby indicating that mass loss due to evaporation is the limiting parameter. A more detailed description is given later in Section 5.

Table 2 Comparison of $H_{YL}$ values (Jurin's Limit) and actual steady state length ($L_c$ ) values for different radii ($R_c$) in the absence of evaporation.

| **Grade** | $R_c$ **(m)** | $t_p$ | $\phi$ | $H_{YL}$ **(mm)** | $L_c$ **(mm)** | $H_{YL}/R_c$ | $L_c/R_c$ |
|---|---|---|---|---|---|---|---|
| 1004 | $75 \times 10^{-6}$ | $205 \times 10^{-6}$ | 0.76 | 74 | 26 | 974 | 347 |
| 1001 | $64 \times 10^{-6}$ | $180 \times 10^{-6}$ | 0.48 | 88 | 15 | 1375 | 236 |
| 1005 | $59 \times 10^{-6}$ | $200 \times 10^{-6}$ | 0.54 | 96 | 12 | 1627 | 210 |

### 3.2 Temperature evolution

We now discuss the temperature evolution in all the cases. Note that being volatile in nature, ethanol is expected to lead to low surface temperatures due to relatively high rates of evaporation (as compared with, say, water). The surface temperature data was obtained through the thermal camera in video mode and the images were extracted through FLUKE SMARTVIEW software.

### a) Transient Analysis:

Figure 5 shows a sequence of extracted thermal images captured at notable time instances. The images are cropped in order to show only the relevant portion, with more clarity and information. In all these images, the temperature increases is followed in the following color order,

(min. temp) Magenta < Blue < Cyan < Green < Yellow < Red (max. temp.)

Two different temperature scales are intentionally shown: global and local. In the global scale, the images (see towards right of the dotted line LL) clearly show the presence of a colder wet (ethanol) zone and hotter dry zone separated by an interface (position of the liquid front). Interestingly, the higher temperature regions (in the wet zone) can be seen at the bottom (from where the liquid ethanol is being pulled out of the relatively 'hotter' reservoir) and near the liquid front. The temperature of the dry zone can be assumed to be at the ambient temperature. In order to get more clarity, we use the local temperature scales where the idea is to not look at the unwanted zone (dry zone in this case). Note that the maximum temperature value in the local scale is 5-6 $^{o}$C lower than that in the global scale. Figure 5a-c show the time sequence of thermal images in local temperature scale for 1001, 1004, and 1005FP, respectively. The respective temperature scales are slightly different across these images while the temperature range is 8-9 $^{o}$C across the three cases.

For a more detailed investigation, we chose to see the surface temperature values at the centreline of these images (marked as 'L0' in Figure 5). Figure 6 shows the height-wise variation of surface temperature along L0 at four different time instants for all the three FPs. length '0' represent the bottom point of the line L0. A common trend in Figure 6 is slowly decreasing temperature values as we move up the line L0 followed by a local temperature minimum at a certain location and then a rapid increase in the surface temperature value eventually leading to a relatively higher constant temperature (equal to the dry surface temperature). A more detailed analysis is required to fully understand this unusual temperature distribution. At t = 120 seconds, the lower temperature regions (wet zone) occupy approximately 8.43mm, 13.26mm, and 6.34 mm for 1001, 1004, and 1005 FP cases, respectively (these values serve only as guideline since the true interface is difficult to track in thermal images with perfection).

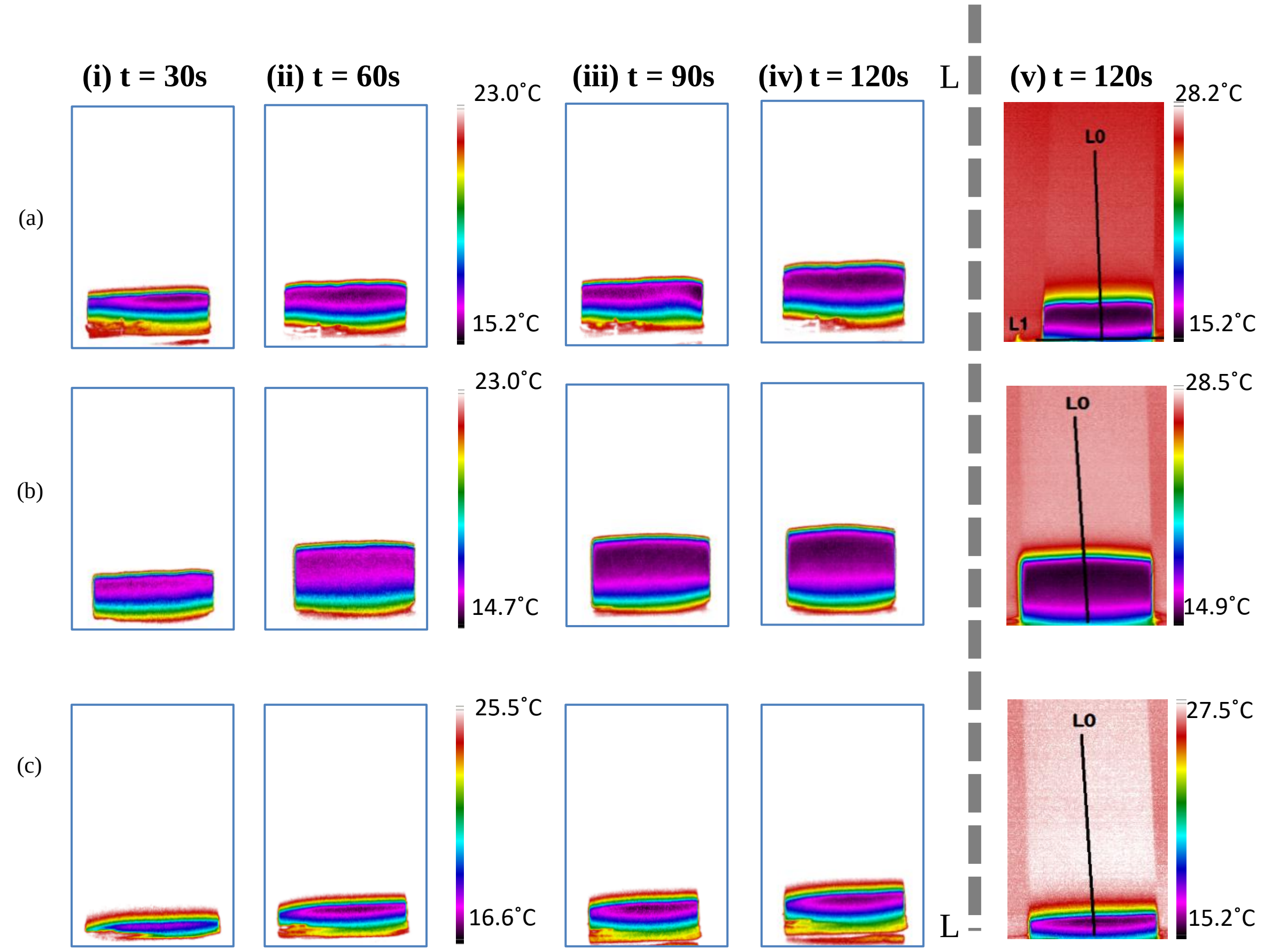

Figure 5 [Colour online] Time sequence of thermal images in local temperature scale for (a) 1001FP, (b) 1004FP, and (c) 1005FP, showing a temperature range of 8-9 ºC. The local temperature scale avoids the dry zone, providing clearer insights into temperature variations. The horizontal line L1 in 'a(v)' represents the reference point from where the liquid wicking length was measured in the FP.

This information is, in fact, directly indicating towards the instantaneous 'h' values. Further, the sudden temperature increase occurs across ~6mm in all the three cases, indicating a significant temperature jump across the liquid front. More discussion on the nature of this temperature curve is given in the next section.

The uncertainty ($U_{T_s}$) [26] in the surface temperature is evaluated as,

$$U_{T_s} = \sqrt{(U_{res})^2 + (U_{sen})^2} \quad (1)$$

Where, $U_{res}$ and $U_{sen}$ are the uncertainty contributed by resolution and sensitivity of thermal camera, respectively. $U_{T_s}$ is estimated to be of the scale of 0.10K, a value closer to the sensitivity of the device.

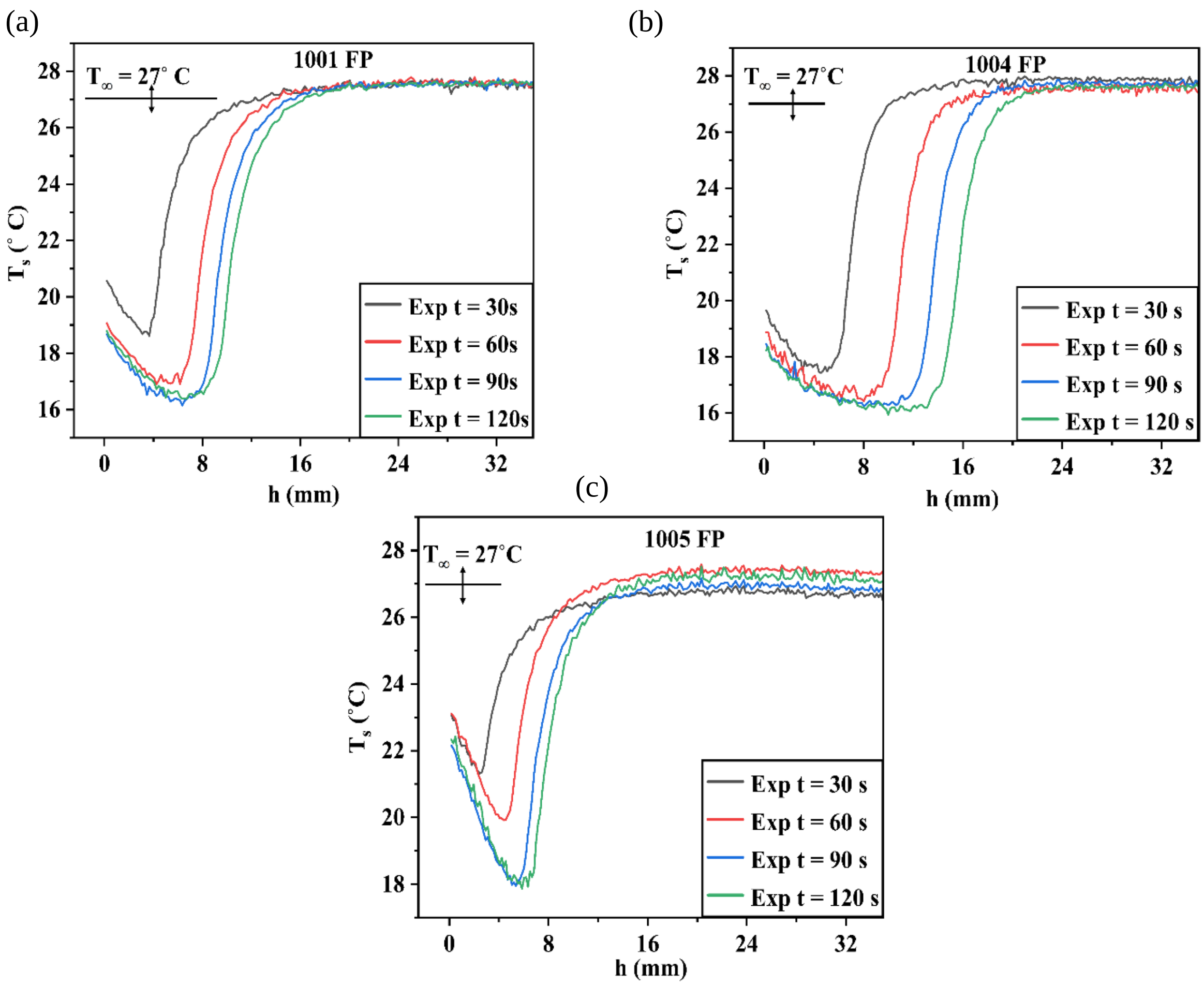

Figure 6 [Colour online] Height-wise variation of surface temperature along the centreline (L0) at four different time instants for (a) 1001FP, (b) 1004FP, and (c) 1005FP. A trend of decreasing temperature followed by a local minimum and a rapid increase is observed, indicating a temperature jump across the liquid front. Double headed arrow in (a) represents the fluctuations (±0.5°C.) in the ambient temperature value throughout the experimental duration.

### b) Steady State Analysis:

This now brings us to discuss the steady state characteristics of the process. Figure 7 shows the thermal images captured in the respective steady state in 1001, 1004, and 1005 FP cases. Note that these images are being seen in the global scale in the temperature range 14-28˚C (this captures all the three cases). Unlike the transient case, a 'warmer' patch is seen, as deep blue colour, between the stable ethanol front and the dry zone seen (similar findings have been reported in [27]). This zone represents the condensed water from the atmospheric water vapour since the temperature reaches below the dew point temperature value. We observed that, in the transient state, condensation did not occur despite temperature values dropping below the dew point. This seems to happen due to two reasons – (a) an already existing liquid surface is a less likely site for nucleation, and (b) the liquid ethanol leading front is in the growth stage. However, in the steady state, the interface may act as a nucleation site as ethanol ceases to rise.

Note that with n-Pentane as the evaporating liquid, the surface temperatures of the FP reach sub-zero and hence the initially condensed water freezes into snow (unpublished).

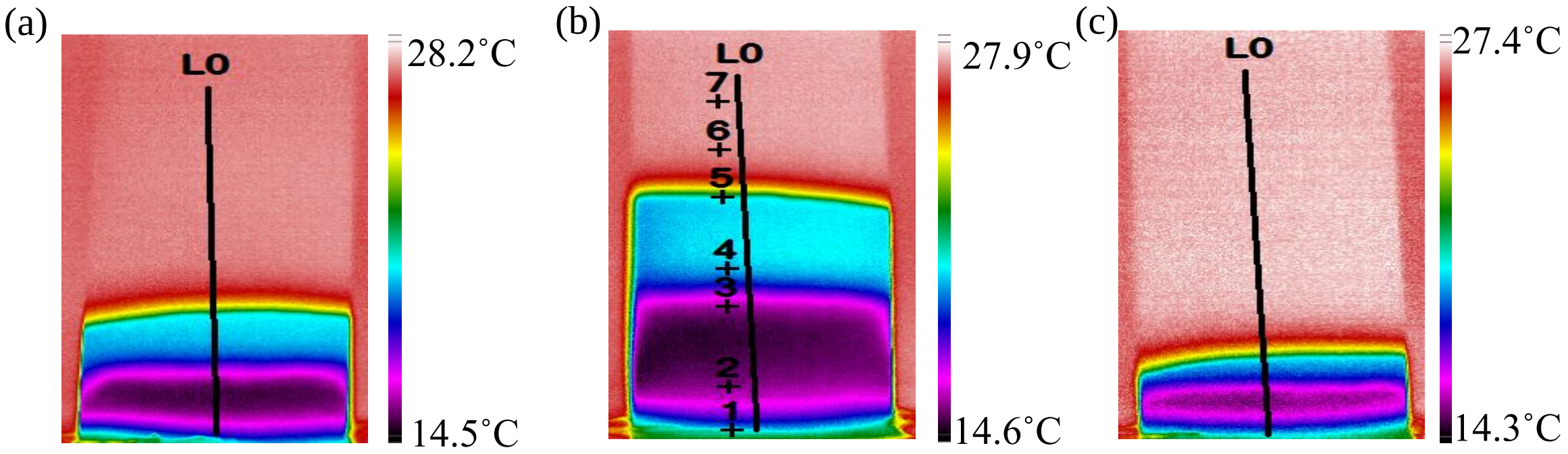

Figure 7 [Colour online] Thermal images in the steady state for (a) 1001FP, (b) 1004FP, and (c) 1005FP, shown in the global temperature scale of 14-28˚C. A warmer patch, indicating condensed water from atmospheric vapor, is observed between the ethanol front and the dry zone.

Similar to the transient cases, we again show the surface temperature variation along the line L0 plotted in Figure 8; which includes all the three cases. In Figure 8a, the temperature time curve for 1004FP has been marked at important points.

The surface temperature reduces quickly in 1-2 path while it is relatively at a constant value in path 2-3. Note that path 1-2-3 represent ethanol (so-called 'wet zone'). Path 4-5 represents the condensed water that can be seen as at a slightly higher temperature than liquid ethanol but at a lower temperature than the ambient. For the 1004FP case, 2-3 and 4-5 in Figure 8a denotes ~15˚C and ~20˚C, respectively, while the ambient (marked as '7' in Figure 8a) is at ~27˚C. Paths 3-4 and 5-6 represent the liquid ethanol- liquid water and liquid water-dry zone interfaces, respectively; they both occupy ~7mm.

Similar conclusion can be drawn for the other cases though the condensed water path is limited in these cases and are not very clearly seen. Figure 8b is a replot of Figure 8a with the key difference being that the vertical axis now denotes the corresponding temperature difference rather than the surface temperature. We see that this difference reaches a maximum value of ~12˚C in all three cases. Another local maxima is seen at ~7˚C and this represents the condensed water. Both these figures have their own utility. This peculiar and rather interesting temperature evolution and height-wise variation indicate that the rate of evaporation is not a constant unlike the one being assumed in Fries et al., 2008 [1] . The significant temperature difference must appear due to varying local evaporation rates. This observation from ours basis for improved mathematical treatment of the entire process. The mathematical treatment begins with a discussion on obtaining permeability value.

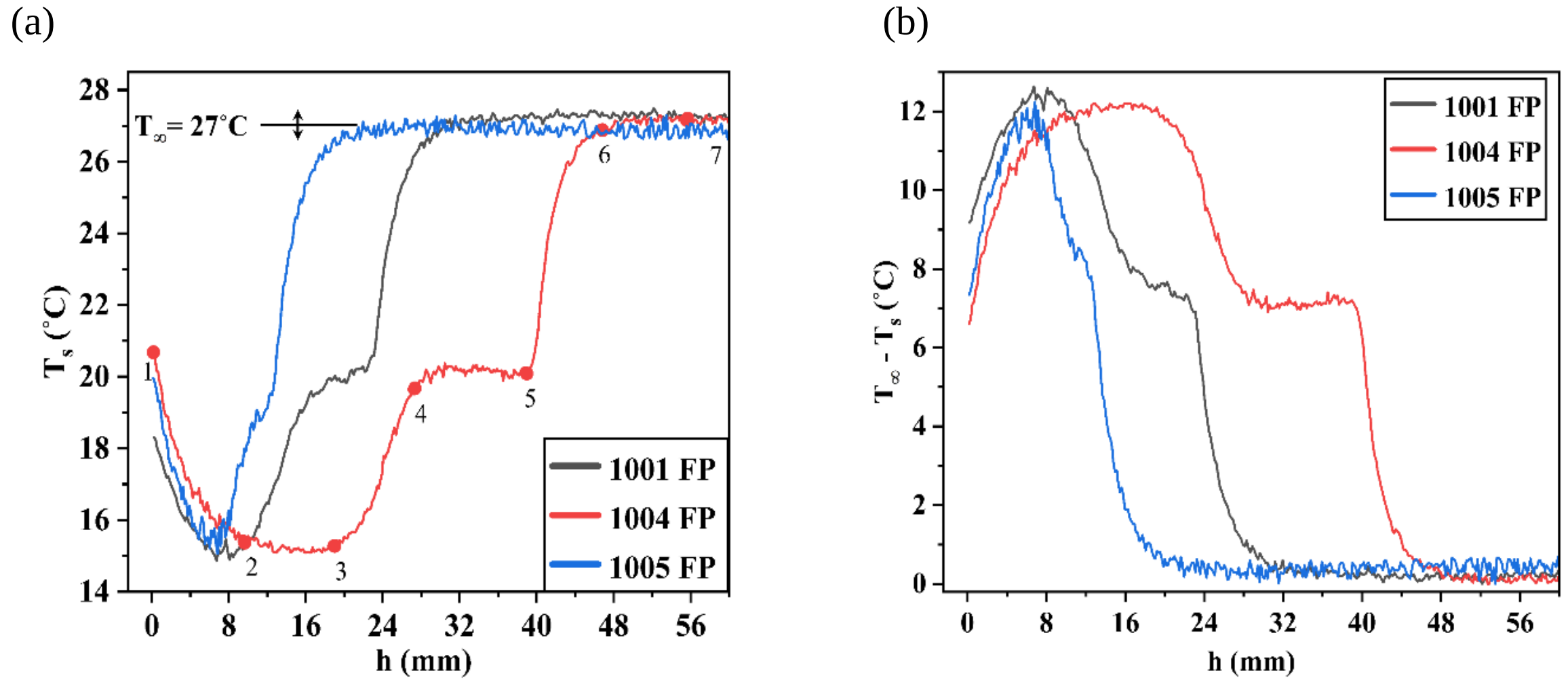

Figure 8 [Colour online] (a) Surface temperature variation along the line L0 for 1001, 1004, and 1005FP in the steady state, showing distinct temperature paths for ethanol and condensed water zones. Double headed arrow in (a) represents the fluctuations ($\pm$0.5℃.) in the ambient temperature value throughout the experimental duration and (b) Replot of Figure 8a with the vertical axis denoting the temperature difference, showing a maximum difference of ~12˚C and a local maxima of ~7˚C representing the condensed water.

## 4. Obtaining Permeability (K) and capillary Radius ($R_c$):

In a non-evaporating vertical system, the surface tension pulls water along the solid surface while the gravitational and viscous forces act against it. The entire process is a competition of these forces leading to the dynamic motion of the liquid front eventually attaining an equilibrium condition.

When a liquid encounters a porous medium, a rapid rise of the liquid occurs. The menisci that drive the wicking are determined by the capillary radius $R_c$. The dynamic pressure balance is given by,

$$\frac{2\sigma\cos\theta}{R_c} = \rho_l g h + \frac{\emptyset}{K}\mu_l h\frac{dh}{dt} \tag{2}$$

The terms in Eq. (2) can be rearranged to give,

$$\frac{a}{h} = b + \frac{dh}{dt} \tag{3a}$$

$$\text{where, } a = \left(\frac{2\sigma\cos\theta}{\emptyset\mu}\right)\frac{K}{R_c} \quad \text{and} \quad b = \left(\frac{\rho_l g}{\emptyset\mu}\right)K \tag{3b}$$

With the known initial condition ($h = 0\ at\ t = 0$), the solution to Eq. (3a) is,

$$t = -\frac{h}{b} - \frac{a}{b^2}\ln\left(1 - \frac{b}{a}h\right). \quad [1] \tag{4}$$

In the initial phase of the experiments (where the evaporation process is not significant), we solved Eq. (4) by feeding the $h$ values and obtained $t$ values by varying the unknowns '$a$' and '$b$' for all three FP cases such that the calculated $t$ values would be nearly equal to the observed $t$ values in the experiments; the consistency was ensured across multiple $t$ and $h$ values in all the cases. From $a$ and $b$ values, we obtained $R_c$ and $K$ values corresponding to each FP case as follows,

$$K = \frac{b\emptyset\mu}{\rho_l g}, R_c = \frac{2\sigma b}{a\rho_l g}, \quad \frac{K}{R_c} = \frac{a\mu\emptyset}{2\sigma} = q_{iter} \tag{5}$$

[28][28][1][9], [29]

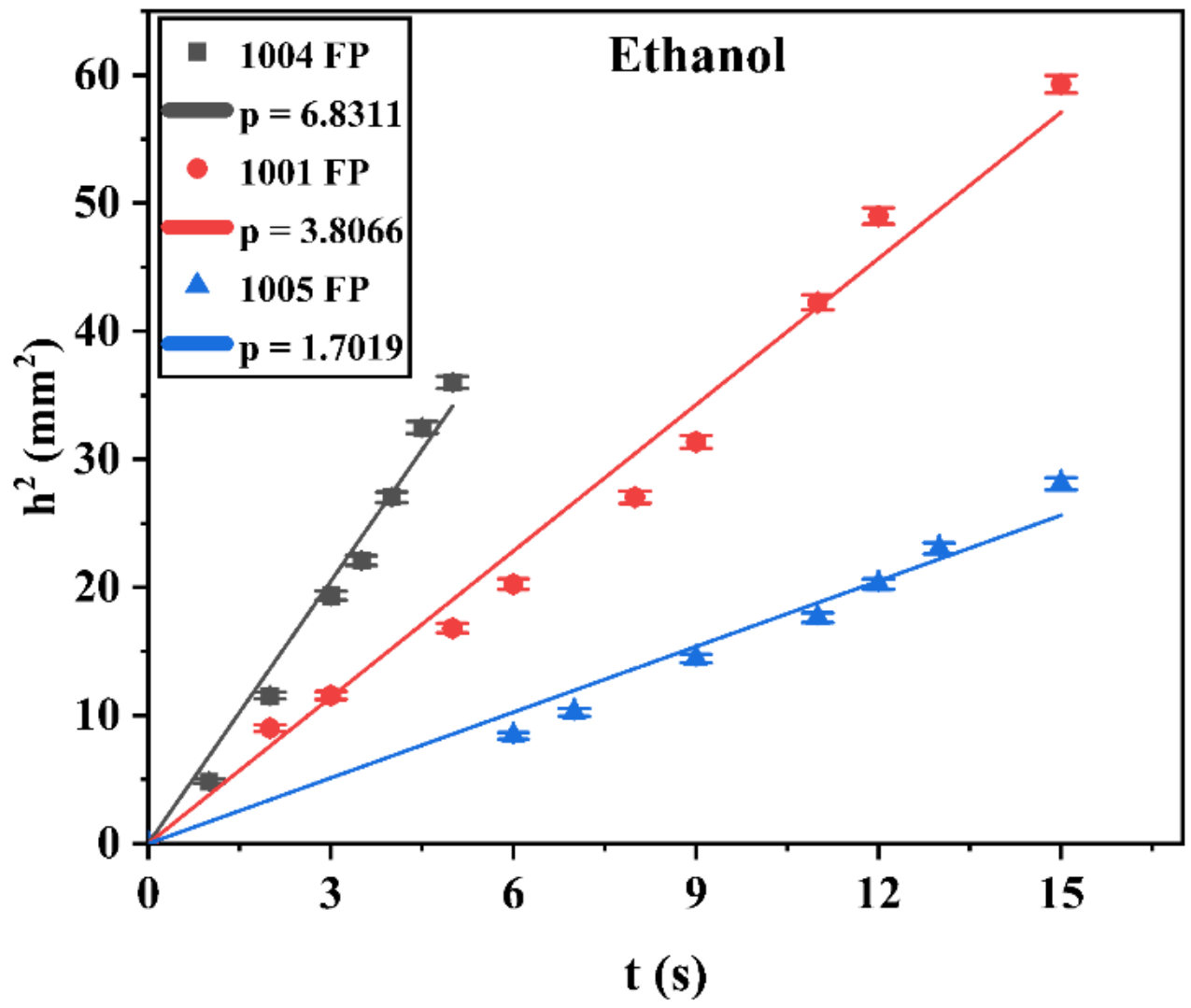

Figure 9 [Colour online] Variation of $h^2$ versus time with linear fits for liquid ethanol wicking on the three FP cases studied here. Values of '$p$' seen in the legend represents the slope of the linear fit in each case with units of $mm^2s^{-1}$. The error bars show the uncertainty $\left(U_{h_i^2}\right)$ in each case.

Eq. (5) yields $R_c$ values to be $59\mu m, 64\mu m, and\ 75\mu m$ for 1005FP, 1001FP, and 1004FP, respectively. Interestingly, when we do not consider the gravity term (which is not a bad consideration in the initial phase of wicking [28]) in Eq. (2), we obtain a linear relationship between $h^2$ and $t$ (the so-called Lucas-Washburn equation [9]). With the initial condition of $h = 0\ at\ t = 0$, the solution is,

$$h^2 = \left[\frac{4\sigma\cos\theta K}{\emptyset\mu_l R_c}\right] t = [p]t \tag{6a}$$

where, $p$ is the slope of $h^2 - t$ experimental curve (see Figure 9). The ratio of $K$ and $R_c$ can be written, hence as,

$$\frac{K}{R_c} = [p]\left(\frac{\emptyset\mu_l}{4\sigma\cos\theta}\right) = q_{exp} \tag{6b}$$

The $q_{iter}$ and $q_{exp}$ values differ only by $\leq 8\%$ across the three FP cases. This ratio can, at times, be tailored depending on the need [30].

The uncertainty $U_{h_i^2}$ [26] for the i-th measurement of the height reached by the liquid front $h_i^2$ is evaluated using the following formula:

$$U_{h_i^2} = \sqrt{\left(2\, h_i\, U_{opt}\right)^2 + \left(\frac{U_d}{2\sqrt{3}}\right)^2} \tag{7}$$

$U_{opt}$ is the uncertainty contribution from the optical resolution as discussed in the Section 3. The uncertainty $U_d$ [31] (depth of dipping) is considered negligible and thus not included in this uncertainty calculation.

Eq. (6a) has been used to plot h$^2$ versus t in our experiments with ethanol, as seen in Figure 9. The slope of the corresponding linear fit yields the value of '$K$'. Note that time data in Figure 9 has been trimmed accordingly such that,

i. In that time period, the temperature difference within the wet zone is small enough so that the evaporation contribution can be ignored and
ii. The liquid has not risen high enough for the wet zone area to become significantly high.

Both the above-mentioned points were ensured in Figure 9 and the time instants turn out to be ~6, ~15, and ~15 seconds for 1004, 1001, and 1005FP, respectively. Given that evaporation effects can be disregarded, these specific time instants roughly represent actual capillary rise. Eq. 6a,b yields the permeability values, and its uncertainty ($U_K$) [26] was calculated as,

$$U_K = \sqrt{\left(\frac{\partial K}{\partial h} U_{opt}\right)^2 + \left(\frac{\partial K}{\partial \emptyset} U_{\emptyset}\right)^2 + \left(\frac{\partial K}{\partial R_c} U_{R_c}\right)^2} \quad (8)$$

In obtaining $U_K$, we did not consider $U_{\emptyset}$ and $U_{R_c}$. The permeability and its uncertainty, hence calculated, are seen in Table 3.

Table 3 Permeability values and their uncertainties based on linear curve fit slope and expanded uncertainty evaluation (Eq. 6 a,b).

| **Liquid** | **Permeability, $K$ ($m^2$)** | | |
|---|---|---|---|
| | **1001 FP** | **1004 FP** | **1005 FP** |
| Ethanol | $1.42 \pm 0.03 \times 10^{-12}$ | $4.86 \pm 0.13 \times 10^{-12}$ | $6.62 \pm 0.05 \times 10^{-13}$ |

Since Eq. (6a) was used only when evaporation is negligible, we analysed the data until which evaporation is small compared to the mass gain rate for ethanol and 1004FP case. Figure 10a shows fairly good agreement between the experimentally obtained rate of mass gain ($\dot{m}_g$) versus time (from experimentally obtained h-t data, $\dot{m}_g = t_p W \emptyset \dot{h} \rho_l$) shown as markers with that obtained by the theory (see Eq. 25 and Eq. 38) shown as solid line. Note that the solution to Eq.25, which is Eq.38, uses the permeability value as well as the evaporation rate value. The initial disagreement between the data seems to be due to the inertial regime being important but ignored in the theoretical formulation.

Figure 10b is an extension of Figure 10a but here only the theoretical solution has been used. At all times, the liquid mass entering the bottom cross-section of the FP is distributed into two components - one used for the wicking (mass gain happening through the cross-section) and the other due to evaporation (mass loss $\dot{m}_l$ occuring mainly from the two large exposed surfaces of the FP, see Figure 12). At any instant, since both these processes are happening simultaneously, the h-t curve obtained either from the experiments or from the theory does not represent the (true) cumulative mass gain rate. We have retained the mass gain rate solution in Figure 10a as the solid line in Figure 10b. The dotted line represents the mass loss rate versus time. This is calculated as the integral of the local mass loss, from an infinitesimally small strip of the FP (with width $W$ and height $dz$), over the instantaneous height of the wet zone. The local evaporated mass employs the use of an average evaporation rate concept that is estimated in the steady state. Since the surface temperature values of the FP is lowest in the steady state, the evaporation rate is the highest. Using this average evaporation rate value in the transient state over-predicts the actual mass loss rate but it gives, to the first approximation, an idea of its magnitude compared to that of mass gain rate. As seen in the Figure 10b, the rate of mass gain for 1004FP dominates the rate of mass loss till ~50 seconds (see inset Figure 10b) and after that at ~100 seconds, both are equal. As mentioned earlier, the cumulative mass gain is the sum of 'mass gain as seen in the FP' and 'mass lost from the FP' and hence it initially follows the mass gain line and at much later time, it follows the mass loss line. This analysis indicates that $t \leq 10$ seconds is an acceptable time window for estimating permeability since evaporative loss has negligible effect on the wicking process. The uncertainty [26] in $\dot{m}_g$ and $\dot{m}_l$ for liquid ethanol wicking on Whatman 1004FP were evaluated as,

$$U_{\dot{m}_g} = \rho_l \, W \, t_p \, \emptyset \, U_{opt} \tag{9}$$

$$U_{\dot{m}_l} = \sqrt{\left(\frac{\partial \dot{m}_l}{\partial \bar{\dot{m}}_e} U_{\bar{\dot{m}}_e}\right)^2 + \left(\frac{\partial \dot{m}_l}{\partial h} U_{opt}\right)^2} \tag{10}$$

Where $U_{\dot{m}_g}$ is uncertainty in mass gain rate and $U_{\dot{m}_l}$, and $U_{\bar{\dot{m}}_e}$ are uncertainties in mass loss rate and height-averaged evaporation rate, respectively. Here $U_{\bar{\dot{m}}_e}$ is dependent on $T_{LV}, \bar{h}, T_\infty$, and $\overline{T_s}$ and evaluated using the uncertainty analysis detailed in Appendix B

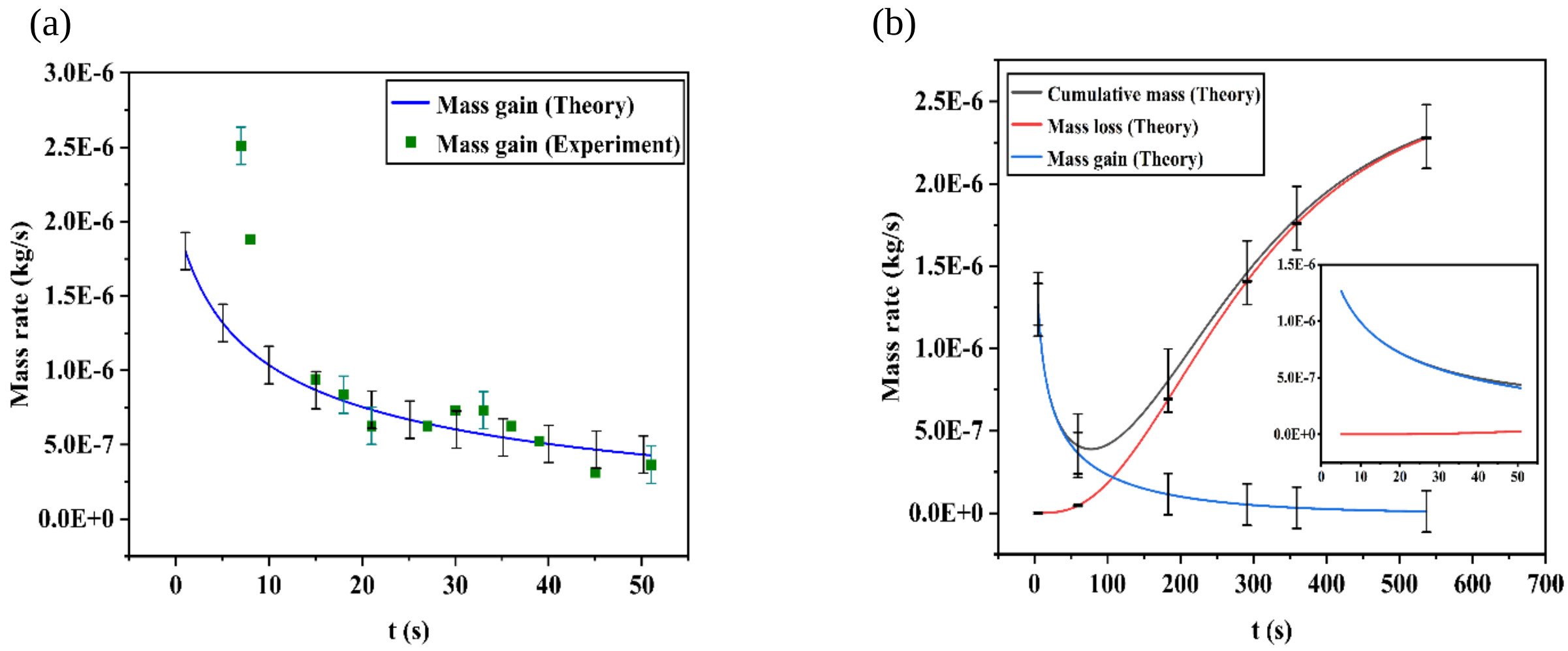

Figure 10 [Colour online] Comparison of experimentally obtained mass gain rate versus time for ethanol and 1004FP with theoretical predictions (solid line), showing good agreement. (b) Extension of Figure 10a showing theoretical solutions; the blue line represents mass gain rate and the red line represents mass loss rate over time, illustrating the dominance of mass gain until ~50 seconds. The error bars represent the respective uncertainties in the corresponding quantities.

## 5. Modification to analytical solution by Fries et al., 2008 [1]

We now describe the entire phenomenon theoretically. As stated earlier, the pioneering work of Lucas-Washburn relates the capillary penetration with time for a horizontal capillary tube. Later, Darcy provided the mathematical framework for flow through a porous medium. These studies did not include evaporation and then the investigation by Fries et al. (2008) [1] did exactly that where the gravitational force was also considered. The pivotal point of Fries analysis was to assume a constant evaporation rate throughout the wet zone. (hereafter named 'constant evaporation model', CEM). In Figure 11 a comparison is shown between Fries et al. 2008 [1] experiment (dotted line) with HFE-7500 and solution to their proposed theoretical model (solid line). The experimental data supports the theoretical model qualitatively, although the theoretical solution model tends to overestimate the reached height by ~35%.

The assumption of a constant evaporation rate seems rather severe in such cases (as mentioned by themselves as well [32]) especially in the current findings where a gradual temperature gradient is observed in the wet zone (see Figure 8a) indicating that the evaporative flux should also vary with height. In this paper, we extend the existing CEM [1], [33] by considering the local evaporation rate ($\dot{m}_e$) as a function of height, '$z$' only (not to vary across the FP width). We explore two different models, one where λ is positive and other where λ is negative as follows,

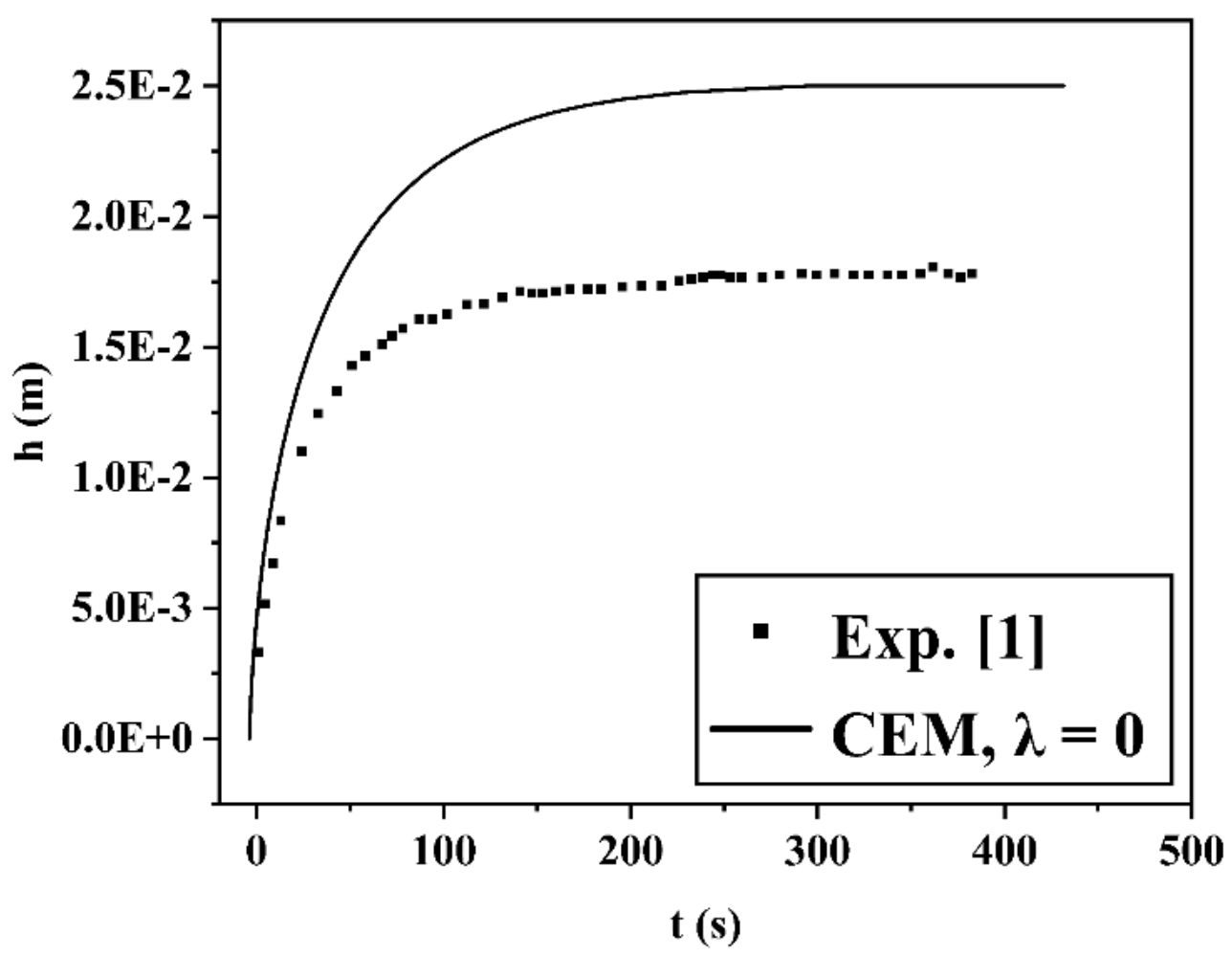

Figure 11 Comparison between Fries et al., 2008 [1] experiment with HFE-7500 (dotted line) and their theoretical model (solid line), showing that the experimental data supports the theoretical model qualitatively, though it overestimates the reached height by ~35%.

$$\dot{m}_e = a_0 \left(\frac{z}{h}\right)^{+\lambda} \qquad + \lambda \; model \tag{11a}$$

$$\dot{m}_e = a_1 \left(1 - \frac{z}{h}\right)^{-\lambda} \qquad - \lambda \; model \tag{11b}$$

$\lambda$ a positive value, is a fitting parameter that quantifies the non-uniformity of the evaporation flux, and $a_0$ and $a_1$ are the corresponding proportionality constants.

***+λ model***

Note that Eq. (11a) satisfies no mass loss at $z = 0$ (see Figure 2d and Figure 12) when $\lambda > 0$ and a constant evaporation rate '$a_0$' at $z = h$.

Note that $\lambda = 0$ in Eq. (11a) returns to the CE model. The detailed solution can be seen in Appendix A. To find the proportionality constant '$a_0$' in Eq. (11a), we integrate the local evaporative mass flow $d\dot{M}(e)$, from an elemental FP strip of total perimeter $2(W + t_p)$ and height '$dz$' from 0 to h,

$$\int_0^h d\dot{M}(e) = \int_0^h -2\dot{m}_e (W + t_p) dz \tag{12}$$

Total evaporative mass flow from the entire FP is,

$$\dot{M}(e) = -2\bar{\dot{m}}_e h (W + t_p) \tag{13}$$

Where $\bar{\dot{m}}_e$ is the height-averaged evaporation rate, and is related to '$a_0$' by

$$\bar{\dot{m}}_e = \frac{a_0}{(1+\lambda)} \tag{14}$$

From Eq. (14) and Eq. (11a), we get

$$\dot{m}_e = \bar{\dot{m}}_e (1 + \lambda) \left(\frac{z}{h}\right)^{\lambda} \tag{15}$$

Note that $\lambda \neq -1$ since it forces $\bar{\dot{m}}_e \rightarrow \infty$, which is unphysical.

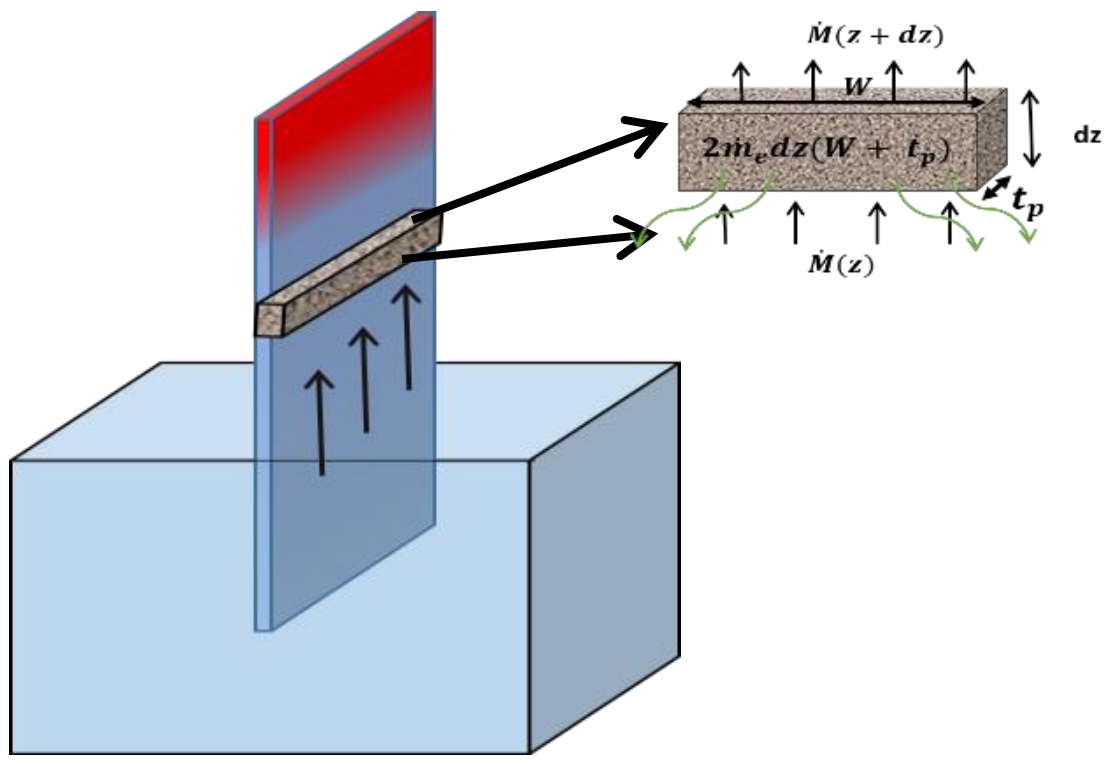

Figure 12 [Colour online] Integral and differential Mass Balance of the FP, showing the total mass inflow $\dot{M}(z = 0)$ comprising mass flow for liquid front movement $\dot{M}_{\dot{h}}$ and total evaporation mass flow $\dot{M}_e$.

The integral and differential mass balance of the FP is shown in Figure 12. It is evident that the total mass inflow $\dot{M}(z = 0)$ consists of two components – the mass flow necessary to supply the movement of the liquid front $\dot{M}_{\dot{h}}$ and the total evaporation mass flow $\dot{M}_e$. The term $\dot{M}_{\dot{h}}$ is given by

$$\dot{M}_{\dot{h}} = \left(\frac{dh}{dt}\right) \rho_l W t_p \emptyset \tag{16}$$

The differential mass balance (see Figure 12) can be expressed as,

$$d\dot{M}(z) = \dot{M}(z + dz) - \dot{M}(z) = -2\dot{m}_e\left(W + t_p\right)dz \tag{17}$$

When integrating and using the boundary condition that the total mass inflow at $z = 0$ must be equal to $\dot{M}_{\dot{h}} + \dot{M}_{\dot{e}}$, one obtains (see Eq. E20 in Appendix A),

$$\dot{M}(z) = \dot{M}_{\dot{h}} + 2\bar{\dot{m}}_e\left(W + t_p\right)\left(1 - \frac{z^{\lambda+1}}{h^{\lambda+1}}\right)h \tag{18}$$

The local mass flow $\dot{M}(z)$ can be expressed as a function of flow velocity, which is used to determine the viscous pressure loss (to be used in the dynamic pressure balance later). Further, flow velocity within the FP can be decomposed into two distinct components. The first one pertains to the velocity of the liquid front, denoted as, $\frac{dh}{dt}$. And, the second component is to replenish the velocity of the system $v_r$; it is necessary to refill the evaporated liquid, which corresponds to its height. It is evident that at z = 0, the velocity of the refill reaches its maximum value.

$$v_{r,0} = \frac{\dot{M}_e}{\rho_l A_b} = \frac{2\bar{\dot{m}}_e h(W + t_p)}{\rho_l W t_p \emptyset} \tag{19}$$

Evaporation varies with height and the refill velocity becomes

$$v_r(z) = v_{r,0}\left[1 - \left(\frac{z}{h}\right)^{\lambda+1}\right] \tag{20}$$

Thus it can be seen that $v_r$ becomes zero at $z = h$ when $\lambda$ is zero. The momentum balance of the liquid inside the FP gives,

$$p_c = p_h + p_{\dot{h}} + p_r \tag{21}$$

Where the individual terms refer to (from left to right):

Capillary pressure, $p_c = \frac{2\sigma \cos\theta}{R_c}$;

Gravity term (hydrostatic pressure, $p_h = \rho_l g h$);

Viscous pressure loss due to $\frac{dh}{dt}$, $p_{\dot{h}}$

Viscous pressure loss due to $v_r(z)$, $p_r$

Viscous pressure loss can be calculated as

$$p_{\dot{h}} = \left(\frac{\emptyset}{K}\mu\right)\int_0^h \dot{h}\, dz = \frac{\emptyset}{K}\mu h\frac{dh}{dt} \tag{22}$$

$$p_r = \frac{\emptyset}{K}\mu \int_0^h v_r(z)\, dz = \frac{\emptyset}{K}\mu v_{r,0} h\left(\frac{\lambda+1}{\lambda+2}\right) \tag{23}$$

Including the varying evaporation rate, the final differential equation becomes

$$\frac{2\sigma\cos\theta}{R_c} = \rho_l g h + \frac{\emptyset}{K}\mu h\frac{dh}{dt} + \frac{\emptyset}{K}\mu\left[\frac{2\bar{m}_e(W+t_p)}{\rho_l W t_p \emptyset}\right] h^2\left(\frac{\lambda+1}{\lambda+2}\right) \tag{24}$$

This can be represented as

$$\frac{a}{h} = b + \frac{dh}{dt} + ch \tag{25}$$

Where the coefficients of a, b, and c are defined as

$$a = \frac{2\sigma\cos\theta}{\emptyset\mu}\frac{K}{R_c} \tag{26}$$

$$b = \frac{\rho_l g K}{\emptyset\mu} \tag{27}$$

$$c = \frac{2\bar{m}_e(W+t_p)}{\rho_l W t_p \emptyset}\left(\frac{\lambda+1}{\lambda+2}\right) \tag{28}$$

Note that $\lambda = 0$ returns $m_e = \bar{m}_e$ (always) and $c = \frac{\bar{m}_e(w+t_p)}{\rho_l w t_p \phi}$ and one obtains CEM solution.

Also if $\bar{m}_e = 0$ and $g = 0$, Eq. (24) returns to a Lucas-Washburn like equation.

Eq. (25) can be rewritten as follows,

$$\int \frac{h}{-ch^2 - bh + a}\, dh = \int 1 dt \tag{29}$$

With the boundary condition of $z = 0$ at $t = 0$, the solution to Eq. (29) is given in the form of $t(h)$ rather than $h(t)$ (see Eq.38) where,

$$\beta = -4ac - b^2 \tag{30}$$

***-λ model***

The same procedure has been followed as mentioned earlier for the $+\lambda$ model, except the form of the local evaporation rate is assumed as,

$$\dot{m}_e = a_1 \left(1 - \frac{z}{h}\right)^{-\lambda} \tag{31}$$

As per Eq. 31. $\dot{m}_e = a_1$ at $z = 0$ and then increase with $z$, which is similar to the nature of $\dot{m}_e$ in $+\lambda$ model as well. The nature of increasing local evaporation rate as '$z$' increases is in line with the reducing surface temperature as '$z$' increases (see Figure 8a). The $-\lambda$ model, however, suggests $\dot{m}_e \rightarrow \infty$ when $z \rightarrow h$ and hence is unphysical in the proximity of '$h$'.

Following the same procedure the final form is,

$$\frac{2\sigma \cos\theta}{R_c} = \rho_l g h + \frac{\emptyset}{K}\mu h \frac{dh}{dt} + \frac{\emptyset}{K}\mu \left[\frac{2\bar{\dot{m}}_e(W + t_p)}{\rho_l W t_p \emptyset}\right]\left(\frac{1}{2-\lambda}\right) h^2 \tag{32}$$

Note that Eq. 32 reduces to CEM when $\lambda = 0$. Further, Eq. 32 can be represented as

$$\frac{a}{h} = b + \frac{dh}{dt} + ch \tag{33}$$

Where the coefficients of a, b and c are defined as

$$a = \frac{2\sigma \cos\theta}{\emptyset\mu}\frac{K}{R_c} \tag{34}$$

$$b = \frac{\rho_l g K}{\emptyset\mu} \tag{35}$$

$$c = \frac{2\bar{\dot{m}}_e(W + t_p)}{\rho_l W t_p \emptyset}\left(\frac{1}{2-\lambda}\right) \tag{36}$$

$$\beta = -4ac - b^2 \tag{37}$$

For $+\lambda$ and $-\lambda$ models, $\beta < 0$ and the final form of the solution is

$$t = \frac{1}{2c}\left[-\ln\left(\frac{-ch^2 - bh + a}{a}\right)\right] - \left[\frac{b}{2c\sqrt{-\beta}} \ln\left\{\frac{(-2ch - b - \sqrt{-\beta})(-b + \sqrt{-\beta})}{(-2ch - b + \sqrt{-\beta})(-b - \sqrt{-\beta})}\right\}\right] \tag{38}$$

From now onwards, Eq. (38) is referred as the solution from the *Non-constant evaporation model* or NCEM.

Note that $-\lambda$ model would probably fail to capture essential physics when $z \rightarrow h$ and $+\lambda$ model would probably fail when $z \rightarrow 0$ (since evaporation rate cannot be zero though evaporated mass loss can be zero due to zero area). Also note that the power law variation of the evaporation rate is only substantial when the surface temperature reduces with increasing $z$ (this is true for a part of the wet zone beyond which the surface temperature increases as seen in Figure 7b and Figure 8a).

We now check the performance of our proposed model in two different ways as mentioned below,

a. Experimental result of Fries et al. (2008) [1] versus CEM versus NCEM and
b. Our present experimental results on 1004, 1001, and 1005 FP versus CEM versus NCEM

In Figure 13, we can clearly see that the experimental results of Fries et al. (2008) [1] did not agree with CEM but agrees fairly well with the $+\lambda$ model when $\lambda = 6$ and $-\lambda$ model when $\lambda = 0.99$.

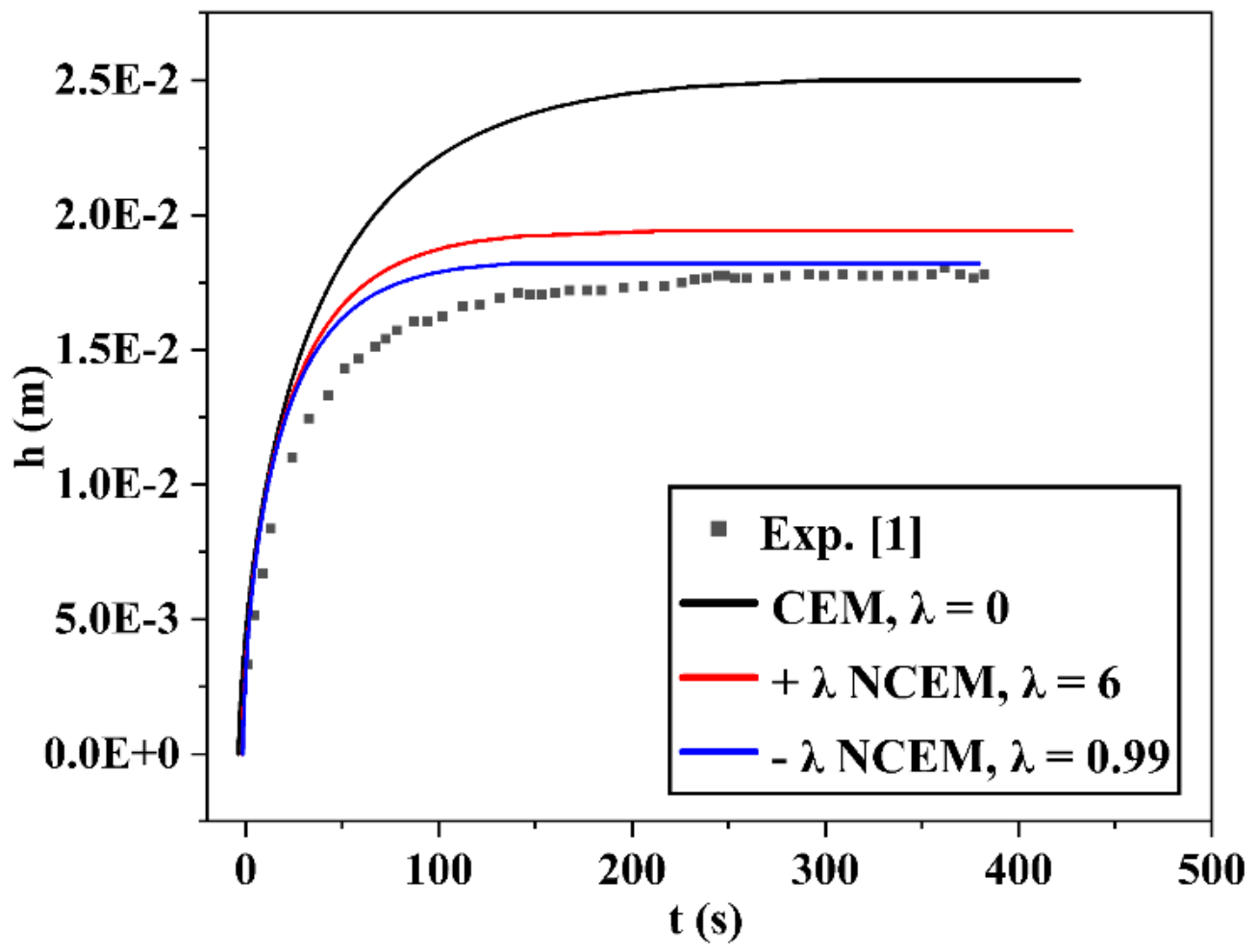

Figure 13 [Colour online] Comparison of experimental results from Fries et al. (2008) [1] with CE model, and NCE model ($+\lambda = 6$, and $-\lambda = 0.99$).

Eq. (38) essentially is a function of parameters ‘a’, ‘b’, and ‘c’ (see Eq. 34-36). In the expressions of ‘a’ and ‘b’, we have obtained ‘$K$’ and hence these are known. The only parameters unknown is $\bar{\dot{m}}_e$ (in Eq. 36). The mass loss measurement through a precision weighing balance does yield $\bar{\dot{m}}_e$. However, we chose to write $\bar{\dot{m}}_e$ explicitly as a function of known parameters some of them are already measured (such as surface temperature). For this purpose we use the surface energy budget (SEB) approach.

## 5.1 Surface Energy Budget:

SEB is a good tool to convert temperature data into mass loss data from an evaporating system [34], [35], [36] and is particularly useful in remote sensing [37]. SEB is essentially a balance of various energy interaction terms between the surface of interest and the surrounding (ambient in this case). Note that the surface temperature of the FP post evaporation of ethanol liquid is lower than that of ambient temperature. The energy required for the evaporation, hence comes from the relatively hotter ambient majority via convection and radiation modes.

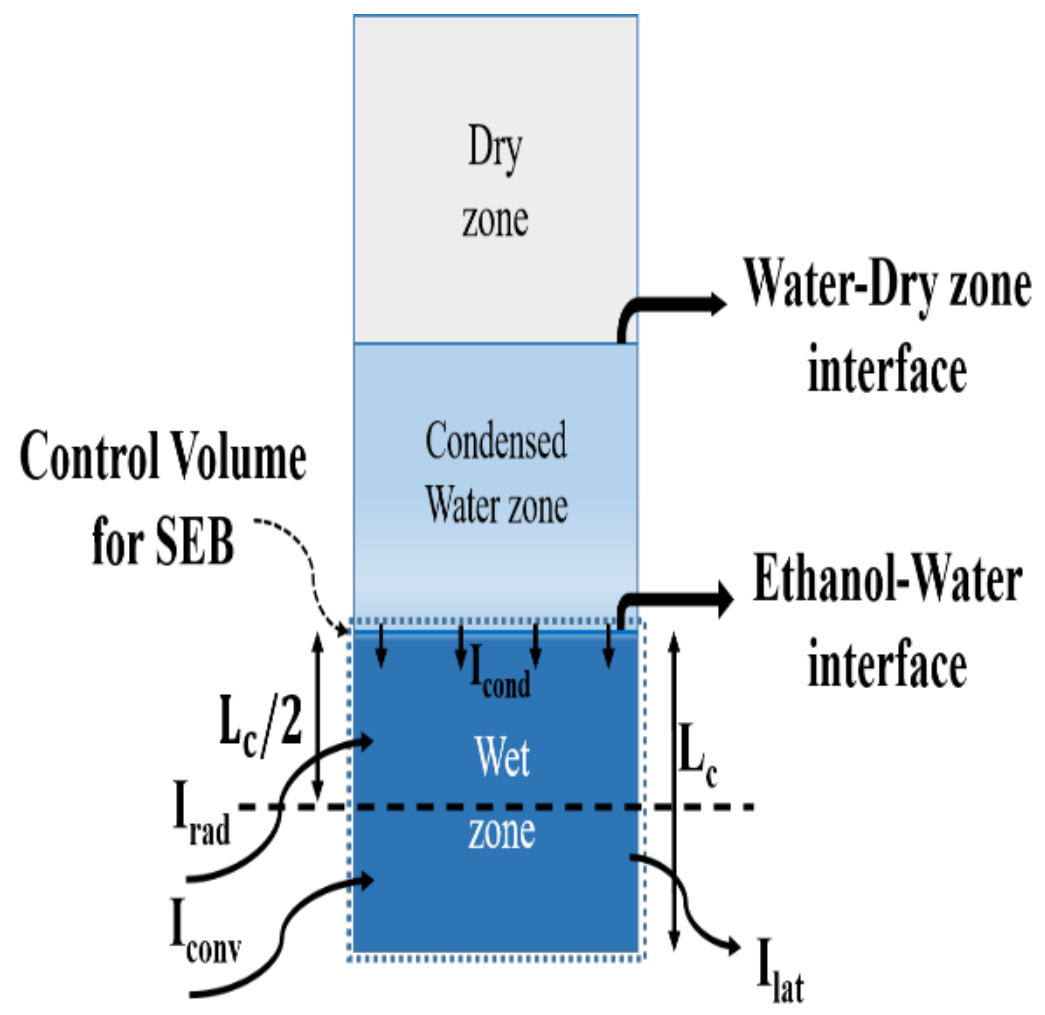

Figure 14 [Colour online] Schematic of the control volume in the steady state for the wet zone. SEB simplifies to energy balance terms between the surface and ambient, primarily through conduction, convection, and radiation.

We consider only the "Wet Zone" marked in Figure 14 as the control volume (CV) for estimating the pixel-wise evaporation rate values (and the evaporation map) through SEB. This was done for two particular reasons: (a) our interest lies mostly on the liquid ethanol wicking and subsequent evaporation dynamics of ethanol and (b) only the liquid ethanol wicking length (or the characteristic length = $L_c$) was used to calculate the governing parameters like Rayleigh Number $\left(Ra_{L_c}\right)$. The algorithm does yield evaporation map for the "Condensed Water Zone" also, but it might not be reliable since the characteristic length of this zone is not $L_c$. Considering the control volume in the steady state, as seen in Figure 14, the SEB is,

$$I_{gain} = I_{loss} \Rightarrow I_{cond} + I_{conv} + I_{rad} = I_{lat} \tag{39}$$

where, $I_{gain}$ and $I_{loss}$ are the heat fluxes received in the CV and lost by the CV, respectively. $I_{lat}$, $I_{cond}$, $I_{conv}$, and $I_{rad}$ are the latent heat loss term due to evaporation from the FP, the heat conducted from the condensed water surface to wet zone, convective heat gain from the ambient, and net radiated heat gain from the ambient, respectively. All the terms in Eq. (39) are in $Wm^{-2}$. Eq. (39) may be written as,

$$\bar{m}_e h_{fg} = \frac{k_L(T_{L-V} - \bar{T}_S)}{\frac{L_C}{2}} + \bar{h}(T_\infty - \bar{T}_s) + \sigma_s \epsilon (T_\infty^4 - \bar{T}_s^4) \tag{40}$$

$$\bar{m}_e = \frac{\frac{k_L(T_{L-V} - \bar{T}_S)}{\frac{L_C}{2}} + \bar{h}(T_\infty - \bar{T}_s) + \sigma_s \epsilon \left(T_\infty^4 - \bar{T}_s^4\right)}{h_{fg}} \tag{41}$$

Where $k_L$ is the thermal conductivity of the liquid, $L_c$ is the steady state length of liquid wicking in the FP. $T_{L-V}$, $\bar{T}_s$, and $T_\infty$ are the temperatures of the liquid-vapor interface, the average surface temperature of the wet zone, and the ambient temperatures, respectively. $h_{fg}$ is the latent heat of vaporization. $\bar{h}$ represents the average convective heat transfer coefficient, which is obtained by the Nusselt-Rayleigh correlation (Churchill and Chu [38]) for the case of natural convection over vertical flat surfaces. Here, the vertical FP is considered as a vertical flat plate. The correlation is expressed as follows:

$$\overline{Nu} = 0.68 + \frac{0.67\, Ra_{L_c}^{1/4}}{\left[1+\left(\frac{0.492}{Pr}\right)^{9/16}\right]^{4/9}}; \qquad Ra_{L_c} \leq 10^9 \tag{42}$$

Rayleigh number is defined as $Ra_{L_c} = \frac{g\Delta\rho L^3}{\bar{\rho}\nu\alpha}$ [39]. $Ra_{L_c}$ is determined by considering the overall change in density of the vapor phase, which is influenced by both the temperature and concentration differences between the evaporated vapors on a wet surface and the ambient, $g$ is the gravitational acceleration, $\nu$ and $\alpha$ are the kinematic viscosity, and thermal diffusivity of air, respectively, and $\Delta\rho = -\rho_\infty + \rho_s$. The density of ambient air is calculated using the following relationship: $\rho_\infty = \frac{p\, m}{k_B\, T}$, where p is absolute pressure (in Pa) calculated at the elevated location of 327 meters (for Jammu, India), m is the molecular mass of dry air (~ 4.81 × $10^{-26}$kg), $k_B$ is the Boltzmann constant (~ 1.380649 × $10^{-23}$ J/K), T is the absolute ambient temperature (K). The liquid-vapor mixture density is calculated using the correlation defined as $ln(\rho_{v,l}) = \alpha_0 + \alpha_0\bar{T}_s + \alpha_2\bar{T}_s^{\,2} + \alpha_3\bar{T}_s^{\,3} + \alpha_4\bar{T}_s^{\,4} + \alpha_5\bar{T}_s^{\,5}$ [40] where $\bar{T}_s$ is surface temperature (in ˚C) air/vapor mixture densities at the liquid surface are calculated as, $\rho_s = \rho_{v,l} + \rho_\infty$.

With the measured average surface temperature of the wet zone and the ambient surface temperature, $Ra_{L_c}$ is calculated, which in turn yields $\overline{Nu}$ from Eq. (42). Note that the experimental $L_c$ value was used in the steady state for this purpose. The obtained $\overline{Nu}$ of the wet zone gives us $\bar{h}$ which is then used in Eq. (41) in order to estimate $\bar{\dot{m}}_e$. Table 4 compares the calculated evaporation rates using SEB, i.e., Eq. (41) to that measured by precision weighing balance.

Table 4 Comparison of calculated evaporation rates using SEB (Eq. 39) with measured evaporation rates by precision weighing balance.

| **Filter Paper Grade** | $\bar{\dot{m}}_e (kg\ m^{-2}\ s^{-1})$ **[SEB]** | $\bar{\dot{m}}_e (kg\ m^{-2}\ s^{-1})$ **[Experimental]** | **% Error** |
|---|---|---|---|
| 1001 | 2.67 ×$10^{-04}$ | 3.08 × $10^{-04}$ | 15.35 |
| 1004 | 2.23 × $10^{-04}$ | 2.77 × $10^{-04}$ | 24.21 |
| 1005 | 2.67 × $10^{-04}$ | 2.69 × $10^{-04}$ | 0.74 |

Since the temperature information can be converted into the evaporation rates using SEB, we now present results of evaporation maps (similar to those obtained remotely [37]). For this purpose, the thermal images were imported in FLUKE SMARTVIEW and pixel wise temperature data was retrieved, which was later post-processed using Eq. (41) to estimate pixel-wise evaporation rate and in turn into the respective evaporation maps. Figure 15d-f displays the evaporation rate maps that corresponds to the respective infrared images in the steady state seen in Figure 15a-c, for 1001, 1004, and 1005 FP cases, respectively. The evaporation map enables us to clearly differentiate between locations of high and low rates of evaporation.

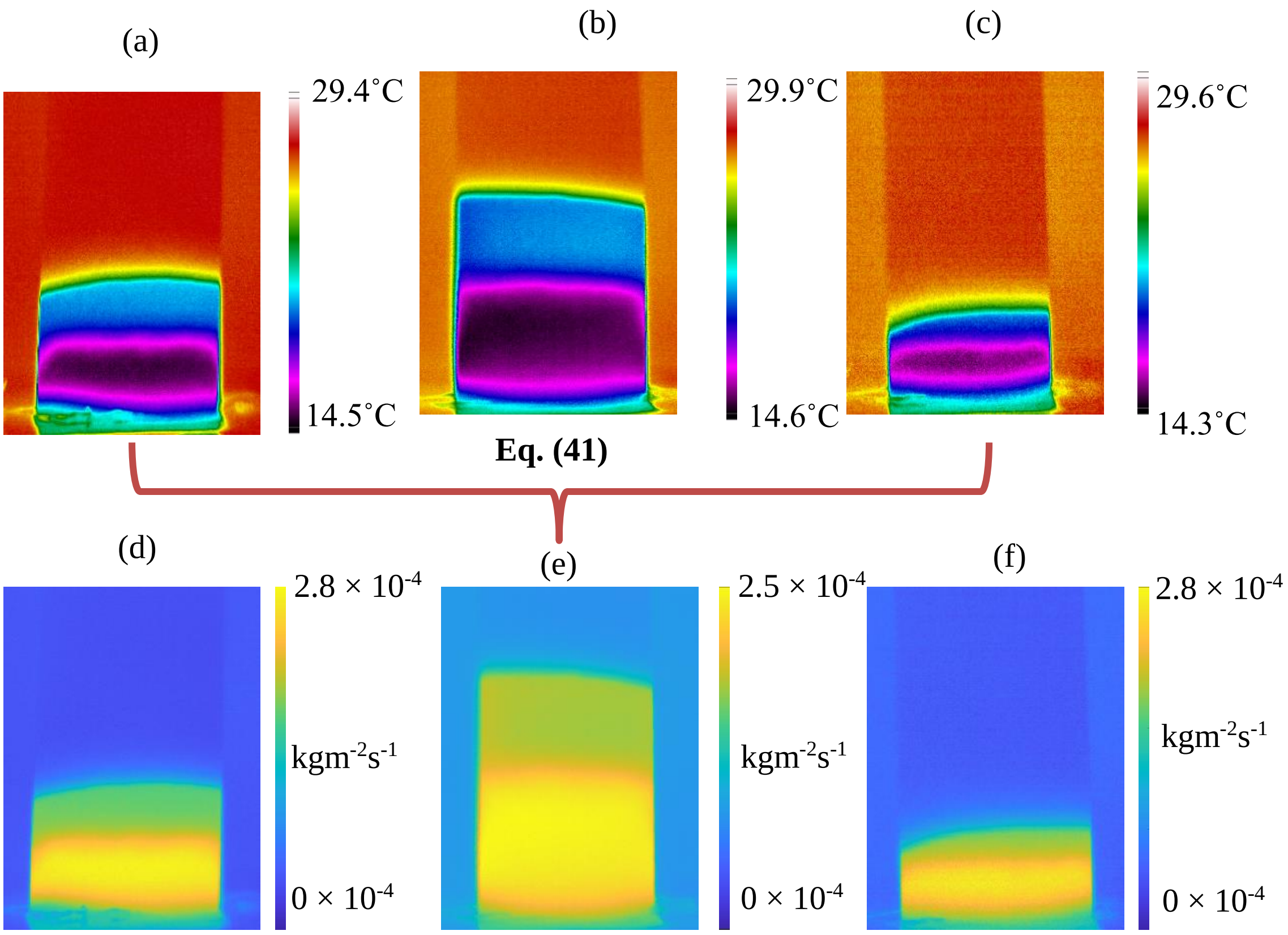

Figure 15 [Colour online] (a-c) show the steady state temperature profiles for Whatman FP's (1001FP, 1004FP, and 1005FP, respectively) and (d-f) show the corresponding evaporation maps ($kgm^{-2}s^{-1}$) at that instant obtained from the SEB. (a,d), (b,e), and (c,f) are correspondingly correlated.

Returning to discussion on the validity of our proposed $+\lambda$ and $-\lambda$ based NCE models, we replace $\overline{\dot{m}}_e$ in Eq. (28) and Eq. (36) by the expression on the R. H. S of Eq. (41). The final expressions for ‘a’, ‘b’, and ‘c’, in the NCE model or Eq. (25) and Eq. (33) are,

$$a = \frac{2\sigma \cos\theta}{\emptyset\mu}\frac{K}{R} \tag{43}$$

$$b = \frac{\rho_l g K}{\emptyset\mu} \tag{44}$$

$$c = \frac{2*\left[\frac{k_L(T_{L-V}-\bar{T}_S)}{\frac{L_C}{2}} + \bar{h}(T_\infty-\bar{T}_S) + \sigma_S\epsilon\left(T_\infty^4-\bar{T}_S^4\right)\right]*(W+t_p)(\lambda+1)}{h_{fg}\rho_l W t_p \emptyset(\lambda+2)} \quad \boldsymbol{+\lambda} \textbf{ model} \tag{45a}$$

$$c = \frac{2*\left[\frac{k_L(T_{L-V}-\bar{T}_S)}{\frac{L_C}{2}} + \bar{h}(T_\infty-\bar{T}_S) + \sigma_S\epsilon\left(T_\infty^4-\bar{T}_S^4\right)\right]*(W+t_p)}{h_{fg}\rho_l W t_p \emptyset(2-\lambda)} \quad \boldsymbol{-\lambda} \textbf{ model} \tag{45b}$$

We used these coefficients in the final updated form of Eq. (38) to obtain $t(h)$ or $h(t)$.

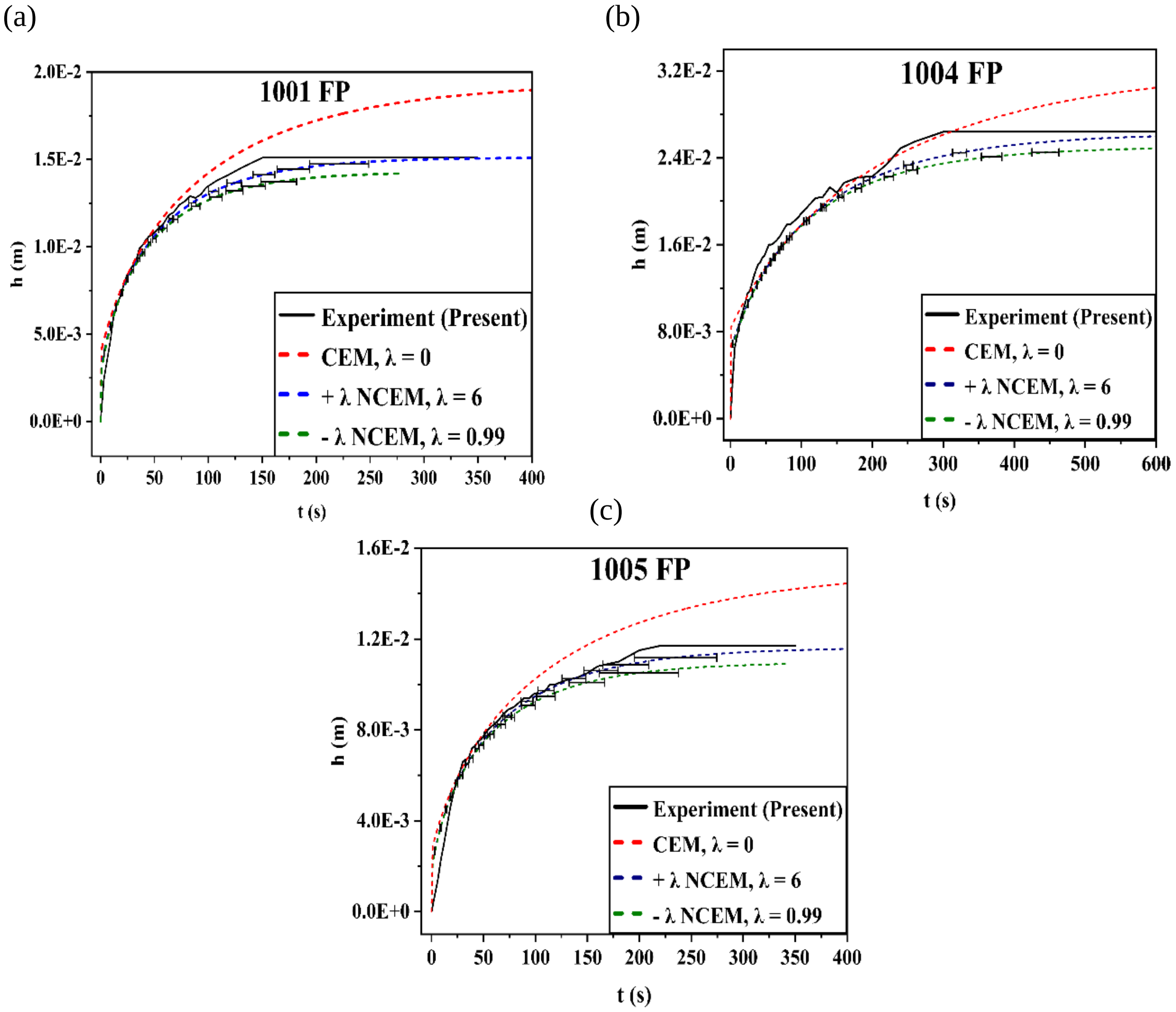

Figure 16 [Colour online] Comparison of the NCE and CE models with experimental data for (a) 1001FP, (b) 1004FP, and (c) 1005FP cases, showing the CE model overestimates while the NCE model aligns well with experiments at +λ = 6 and -λ = 0.99 across all FP cases. The horizontal error bars represent the total uncertainty in time calculated as per Eq. (38), see Appendix B.

Now we compare the current model NCE model with that of CE model for the experimental data in 1001FP, 1004FP, and 1005FP cases. Figure 16 shows such a comparison in all the three cases, the CE model overestimated the experimentally observed h-t curves. However, the proposed NCE model agrees reasonably well with the experiments when $\lambda = 6$ for $+\lambda$ model and $\lambda = 0.99$ for $-\lambda$ models in all the three cases studied here. Interestingly, these $\lambda$ values are observed to be consistent across the three cases.

## 5.2 Severity analysis of the proposed NCE model

We now look at the severity of the proposed NCE model. For this purpose, we plot $\dot{m}_e$ versus $z$ in Figure 17 for all the three cases. Note that the horizontal dotted lines in Figure 17 represent constant evaporation rates corresponding to the respective cases and are obtained from CE model. The dashed and solid lines represent solution from $-\lambda$ and $+\lambda$ NCE models, respectively. The black circles represents local evaporation rates obtained from the SEB. The area under the $\dot{m}_e - h$ curve for a particular case is constant irrespective of the model used. In all the three cases, evaporation rate from $-\lambda$ NCE model shoots up suddenly at ~95% of the respective $L_c$ values. Note that this model yields $\dot{m}_e \to \infty$ when $z \to L_c$ and hence this model may not be trust worthy at heights closer to that of the respective steady state penetration lengths. On the contrary, $+\lambda$ NCE model predicts a non-linear increase in the evaporation rates quite sooner (~50% of $L_c$ values). At ~75% of $L_c$ value, the $\dot{m}_e$ value from $+\lambda$ NCE model reaches the constant $\dot{m}_e$ value from CE model. Afterwards, these values are higher till ~$L_c$ as can be seen in Figure 17. $+\lambda$ NCE model successfully captures majority of the physical phenomena undertaken in this problem. However, this will also suffer at higher '$z$' values since the evaporation rates are expected to reach a maximum corresponding to the respective minima (see Figure 8a) and decreased values afterwards. Note that the local evaporation rates from SEB vary significantly (30% being the highest and ±15% from the average evaporation rate). This variation shows that the constant evaporation rate value (or model) is insufficient to capture the actual dynamics. Even $-\lambda$ and $+\lambda$ NCE models do not predict the actual evaporation rates (and their variation), but they capture the $h - t$ curve quite well. In future communications, we intend to incorporate a more suitable form of the height-wise variation of the evaporation rate keeping the merit of h-t curves intact while also attempting to answer this fascinating evaporation rate trend that might be an outcome of the interplay between the conduction heat gain from the dry zone and the nature of liquid vapour transport (heavier than air) governed primarily by the concentration boundary layer (along with the thermal boundary layer).

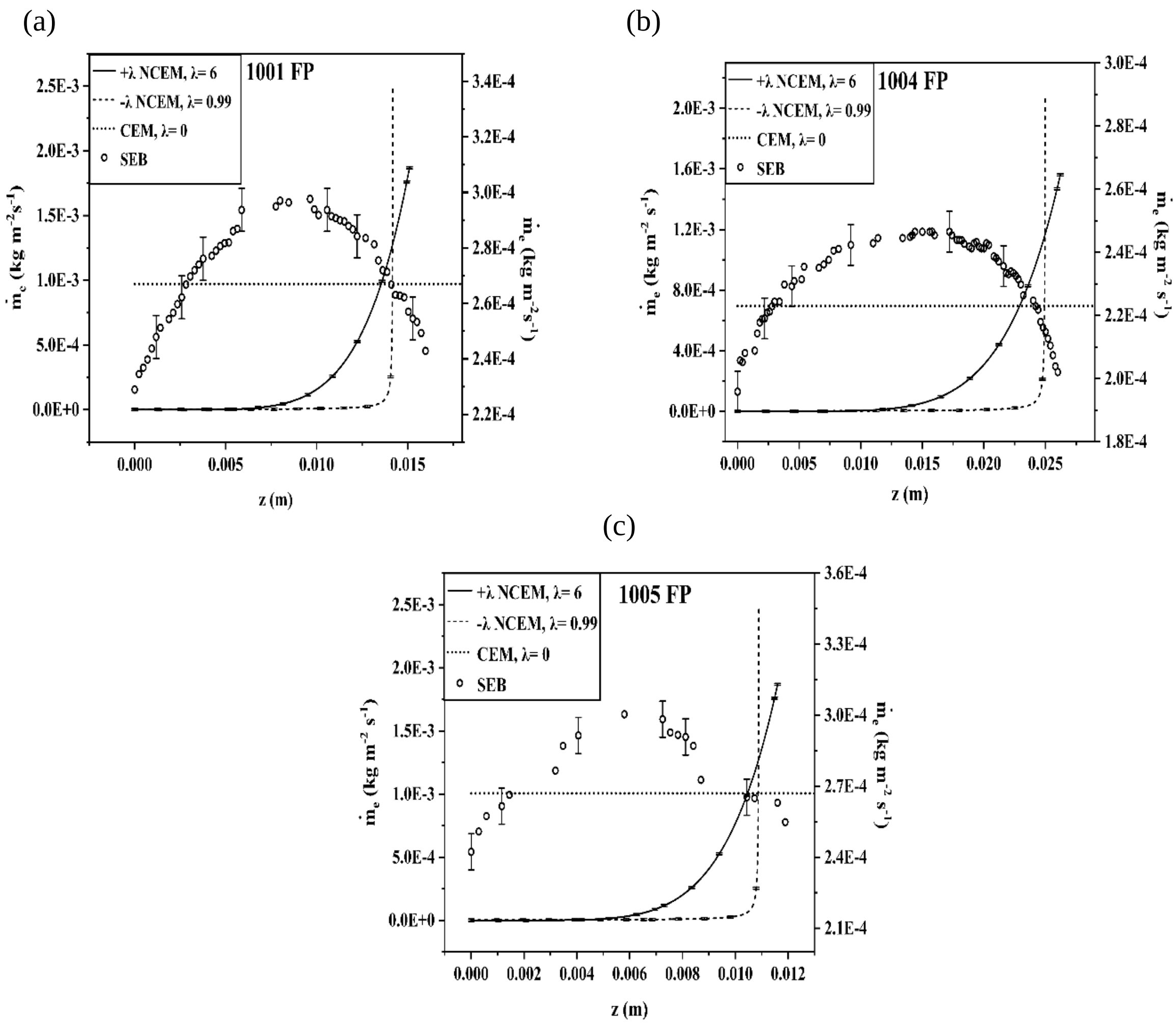

Figure 17 Evaporation rate $\dot{m}_e$ versus height *z* for all three cases. Black circle shows the $\dot{m}_e$ data from SEB, short dot show constant rates from the CE [1] model which have to be read on the secondary y-axis, while short dashed (-λ) and solid (+λ) lines indicate solutions from NCE models to be read on the primary y-axis,[1] illustrating distinct behaviours nearing respective $L_c$ values with (a) 1001 FP, (b) 1004 FP, and (c) 1005 FP. The error bars represent the uncertainty in evaporation rates ($U_{\dot{m}_e}$).

The uncertainty [26] in the calculated rates of evaporation (see Eq. 15) can be written as,

$$U_{\dot{m}_e} = \sqrt{\left(\frac{\partial \dot{m}_e}{\partial \bar{\dot{m}}_e} U_{\bar{\dot{m}}_e}\right)^2} \quad (46)$$

Here, $\bar{\dot{m}}_e$ is dependent on $T_{LV}, \bar{h}, T_\infty$, and $\bar{T}_s$ (see Eq. 41). The procedure follows the methodology discussed in Appendix B, where individual uncertainties are mentioned in Table 5.

### 5.3 Time scales & Dimensional analysis

We now look at the relevant scales or order of magnitude of time and height involved in the entire wicking process with evaporation. Fries et al., 2008 [1] did mention the relevant dimensionless numbers, the so-called ‘Height Number (HN)’ and ‘Time Number (TN)’. Note that HN and TN, both were devised in a way to involve the competition between the surface tension force, gravitational force, and the viscous force while not involving the mass loss related term. In a process where evaporation is the major limiting factor, mass loss aspect seems to be included straight forward in the scaling arguments. Further, note that the entire experimental process reaches steady state quite fast (~200 seconds for ethanol with 1001, 1004, and 1005 FP in the present cases and ~100 seconds for 'High Evaporation' case in Fries et al. 2008 [1]) and we believe that this time duration is probably insufficient for gravitational forces to become significant, while the viscous forces do play a role (see Section 4). This indicates that the balance is likely to happen between surface tension, viscous, and mass loss related terms.

The height and time scales based on the previous arguments are,

$$\text{Height scale} \sim \sqrt{\frac{a}{c}} = \sqrt{\frac{\sigma K \rho_l t_p}{\bar{\dot{m}}_e \mu R_c}} \quad (47)$$

$$\text{Time scale} \sim \frac{1}{c} = \frac{\rho_l t_p}{\bar{\dot{m}}_e} \quad (48)$$

The 'height number' and 'time number' used by Fries et al. 2008 [1] are mentioned below along with our proposed 'Evaporation Height Number (EHN)' and 'Evaporation Time Number (ETN)';

$$\text{HN} = \frac{hb}{a} = \frac{hR_c\rho_l g}{2\sigma\cos\theta} \qquad \text{TN} = \frac{tb^2}{a} = \frac{tR_c\rho_l{}^2 g^2 K}{2\emptyset\sigma\mu\cos\theta} \quad (49)$$

$$\text{EHN} = z\sqrt{\frac{c}{a}} = z\sqrt{\frac{\bar{\dot{m}}_e \mu R_c}{\sigma\rho_l K t_p}} \qquad \text{ETN} = ct = \frac{t\,\bar{\dot{m}}_e}{\rho_l t_p} \quad (50)$$

EHN and ETN are based on the competition between mass gain factor (by surface tension) and mass loss factor (i.e., Evaporation); we do not consider the viscous and gravitational forces as they do not seem to effect the process significantly. Note that the average evaporation rate in the steady states was used to obtain these scales but in principle the temporal values can also be used for scaling.

Figure 18a shows the variation of HN versus TN (by Fries et al. 2008 [1]); the inset shows the variation of same data plotted on a logarithmic scale in the vertical axis). The curves do collapse up to a vary small value of TN (~0.01) and deviate beyond it eventually settling at totally different HN values in the respective steady states. The choice of HN and TN, hence, do not seem appropriate for such a process. [1]Figure 18b shows the variation of EHN versus ETN following Eq. (50). The EHN values for all the cases (including Fries et al., 2008 [1]) are ~1 in the steady state which arrive at ETN ~0.35. All the four curves in Figure 18 seem to follow a trend and besides their different $L_c$ values, their steady state EHN values are quite close, thereby, indicating that EHN and ETN might be better choices to signify the dimensionless behavior of a volatile liquid climbing on a porous medium. For a combination of a porous medium and an evaporating liquid, these EHN and ETN values can be directly used to estimate the steady state time and lengths ($L_c$). For example: when we consider the 'high evaporation' case in Fries et al. 2008 [1], an ETN value of 0.35 yields the steady state time of 194 seconds and an EHN value of 1.0 yields an $L_c$ value of 17.9mm. The experimentally measured steady state length was ~18.0mm (Figure 11) that was achieved in ~250 seconds (steady state time). This totally different combination of a porous medium and a liquid (HFE 7500 in Fries et al., 2008 [1]) yielded values very close to those from EHN and ETN in the steady state showing the relevance and significance of the two evaporation-based dimensionless numbers.

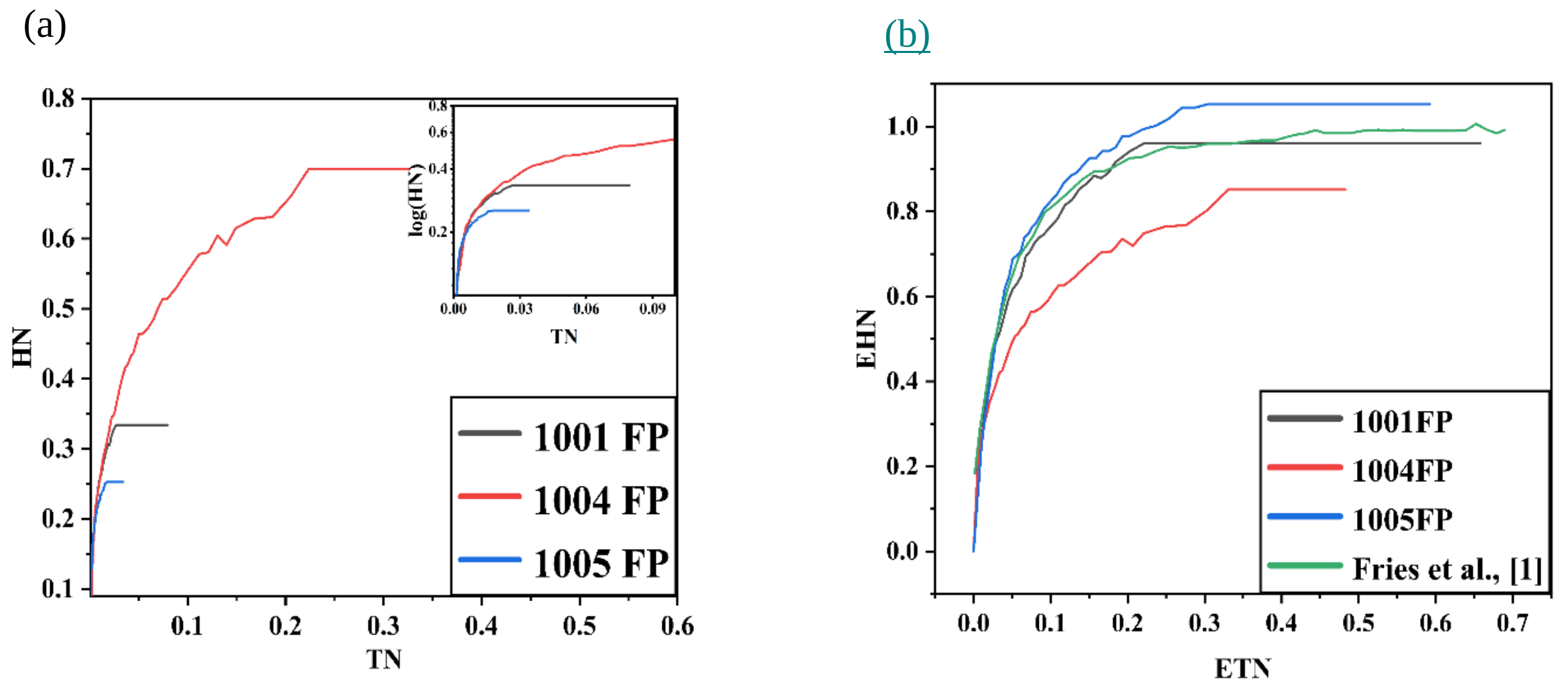

Figure 18 [Colour online] (a) Variation of HN versus TN (by Fries et al. 2008 [1]) and [1](b) variation of EHN versus ETN, indicating ETN as a potentially more suitable time scale in cases dominated by evaporation, with EHN values converging to ~1 in steady state.

## 6. Conclusions

We report findings on the investigation of the process of wicking of ethanol on vertical rectangular porous strips (filter papers). The experiments were conducted, in controlled laboratory conditions, on three different FPs of different permeabilities, and the intention was to understand the importance of evaporation on the liquid rise dynamics. For this purpose, three diagnostic tools were used simultaneously, i.e., optical imaging, thermal imaging, and mass data from a precision weighing scale. The liquid rises in the vertically oriented FPs with time and reaches a particular steady state height ($L_c$) that is much lower compared to Jurin's limit. The findings, therefore, indicate that evaporative mass loss is the primary limiting factor in such (volatile) cases rather than the gravitational forces (in conventional systems).

The thermal images supplemented the optical data and also revealed a very interesting temperature distribution along the FP. The lower surface temperature values were evident due to the cooling effect caused by evaporation but the resulting temperature gradient showed that the evaporation rate of liquid is not constant along the vertical direction (a significant deviation from the previously reported work by Fries et al., 2008[1]). Probably, this non-constant temperature variation led to a $\sim35\%$ difference between the CEM and experimental value in their research. Accordingly, we proposed two different models (namely, non-constant evaporation model, NCEM) where power law variation of the rate of evaporation was inputted in the model. Capillary radius $R_c$ and permeability value $K$ were estimated by fitting the experimental data to the theoretical model given by Fries et al., 2008[1] (for non-evaporating cases, that seems to be applicable to the evaporating cases as well in the initial time instants when the evaporative mass loss is not significant) while ensuring consistency across multiple $t$ and $h$ values that were experimentally measured. The SEB was used to write the local evaporation rate as a function of measured temperature value and this was plugged into the momentum equation that eventually yield t(h). $+\lambda$ NCE model with $\lambda = 6$ resulted in a better agreement (within $\pm1\%$) with those from the experiments. However, a better model should be able to capture the temperature inversion (see Figure 8a) as well and this remains a work for the future.

The time duration to reach steady state height ($L_c$) clearly suggests that the balance is majorly between the surface tension force, viscous force, and viscous force (generated by evaporation-induced refill velocity). The system can, hence, be seen also as a balance of masses - mass gain due to surface tension and mass loss due to evaporation. When these two are equal steady state is achieved. This balance yields EHN and ETN and seem to be the better non-dimensional numbers than the known HN and TN by Fries et al., 2008 [1]. In fact EHN versus ETN data collapsed for totally different combinations of porous media and liquids (FP and ethanol in the present case while Fries et al., 2008 [1], used a metal wick and HFE 7500). The proposed EHN and ETN involve evaporation rate term that acts as the major limiting parameter. The EHN and ETN formulations can be considered as a step towards devising a universal scaling in such processes. This study is useful not only in terms of a better understanding of such a well-studied process but also in thermal management of wicks and other cooling devices like heat pipes, etc.

**Acknowledgements**

Rampally Srirama Chandra Murthy did conceptualization, numerical implementation of the methods, analysis of the results, and writing of the original manuscript. Navneet Kumar worked on conceptualization, supervision, analysis of the results, and reviewing and editing of the manuscript. We thank the anonymous referees for their valuable suggestions.

**Statements & Declarations**

**Funding**

The authors greatly acknowledge the financial support provided under SEED grant by IIT Jammu and SRG/2022/000680 by SERB, DST (GoI).

**Competing interests**: The authors report no competing interests. The authors alone are responsible for the content and writing of the paper.

**Data Availability**

The datasets generated during and/or analysed during the current study are available from the corresponding author on reasonable request.

**Corresponding author**

Correspondence to Rampally Srirama Chandra Murthy

**Appendix A: The impact of varying evaporation rate in NCEM models**

***+λ model***

From Figure 12, we can write differential mass balance as,

$$\dot{M}(z) = \dot{M}\,(z + dz - dz) \tag{E1}$$

$$\Rightarrow \dot{M}(z) = \dot{M}\,(z + dz) - d\dot{M}(z) \tag{E2}$$

$$\Rightarrow d\dot{M}(z) = -2\dot{m}_e\left(W + t_p\right)dz \tag{E3}$$

Here, the evaporation rate $\dot{m}_e$ is a function of z, then we can write it as,

$$\dot{m}_e = a_0\left(\frac{z}{h}\right)^{\lambda} \tag{E4}$$

To find the evaporation constant ‘$a_0$’, we integrate from 0 to h. We get

$$\int_0^h d\dot{M}(e) = \int_0^h -2\dot{m}_e\left(W + t_p\right)dz \tag{E5}$$

Total evaporative mass flow, $\dot{M}(e) = 2\overline{\dot{m}}_e\left(W + t_p\right)h$ (E6)

$$\Rightarrow \dot{M}(e) = \int_0^h -2\dot{m}_e\left(W + t_p\right)dz \tag{E7}$$

$$\Rightarrow \dot{M}(e) = \int_0^h -2a_0\left(\frac{z}{h}\right)^{\lambda}\left(W + t_p\right)dz \tag{E8}$$

From Eq. (E6) and Eq. (E8), we have

$$-2\bar{\dot{m}}_e\left(W + t_p\right)h = -2a_0\frac{(W+t_p)}{(\lambda+1)}h \quad \text{(E9)}$$

$$\Rightarrow \bar{\dot{m}}_e = \frac{a_0}{(\lambda+1)} \quad \text{(E10)}$$

We substitute $a_0$ in Eq. E4 to get $\dot{m}_e = \bar{\dot{m}}_e(\lambda+1)\left(\frac{z}{h}\right)^{\lambda}$ (E11)

When integrating and using the boundary condition that the total mass inflow at z = 0 must be equal to $\dot{M}_{\dot{h}} + \dot{M}_{\dot{e}}$, one obtains,

$$\int d\dot{M}(z) = \int -2\dot{m}_e\left(W + t_p\right)dz \quad \text{(E12)}$$

$$\Rightarrow \dot{M}(z) = \int -2\bar{\dot{m}}_e(\lambda+1)\left(\frac{z}{h}\right)^{\lambda}\left(W + t_p\right)dz \quad \text{(E13)}$$

$$\Rightarrow \dot{M}(z) = -2\bar{\dot{m}}_e(\lambda+1)\frac{(W+t_p)}{(\lambda+1)}\, z\left(\frac{z}{h}\right)^{\lambda} + c \quad \text{(E14)}$$

$$\Rightarrow \dot{M}(z) = -2\bar{\dot{m}}_e\left(W + t_p\right) z\left(\frac{z}{h}\right)^{\lambda} + c \quad \text{(E15)}$$

Applying the boundary condition that at $z = 0$, $\dot{M}(z) = \dot{M}_{\dot{h}} + \dot{M}(e)$

$$\dot{M}_{\dot{h}} + \dot{M}(e) = -2\bar{\dot{m}}_e\left(W + t_p\right) 0\left(\frac{0}{h}\right)^{\lambda} + c \quad \text{(E16)}$$

$$\Rightarrow c = \dot{M}_{\dot{h}} + \dot{M}(e) \quad \text{(E17)}$$

Substitute ‘c’ in Eq. (15) we get,

$$\dot{M}(z) = -2\bar{\dot{m}}_e\left(W + t_p\right) z\left(\frac{z}{h}\right)^{\lambda} + \dot{M}_{\dot{h}} + \dot{M}(e) \quad \text{(E18)}$$

We know that $\dot{M}(e) = 2\bar{\dot{m}}_e\left(W + t_p\right)h$ (E19)

$$\dot{M}(z) = -2\bar{\dot{m}}_e\left(W + t_p\right) z\left(\frac{z}{h}\right)^{\lambda} + \dot{M}_{\dot{h}} + 2\bar{\dot{m}}_e\left(W + t_p\right)h \quad \text{(E20)}$$

$$\Rightarrow \dot{M}(z) = \dot{M}_{\dot{h}} + 2\bar{\dot{m}}_e\left(W + t_p\right)h\left[1 - \frac{z^{\lambda+1}}{h^{\lambda+1}}\right] \quad \text{(E21)}$$

The local mass flow $\dot{M}(z)$ can be expressed as a function of flow velocity, which is subsequently utilized to determine the viscous pressure loss.

At the point where z equals zero, the velocity of the refill reaches its maximum value.

$$v_{r,0} = \frac{\dot{M}_e}{\rho_l A_b} = \frac{2\bar{\dot{m}}_e h(w+t_p)}{\rho_l w t_p \emptyset} \quad \text{(E22)}$$

Evaporation rate varies with height and the refill velocity becomes,

$$v_r(z) = v_{r,0}\left[1 - \left(\frac{z}{h}\right)^{\lambda+1}\right] \quad \text{(E23)}$$

The momentum balance of the liquid inside the FP is,

$$p_c = p_h + p_{\dot{h}} + p_r \quad \text{(E24)}$$

Viscous pressure loss can be calculated as,

$$p_{\dot{h}} = \frac{\emptyset}{K}\mu \int_0^h h\,dz = \frac{\emptyset}{K}\mu h\dot{h} \tag{E25}$$

$$\Rightarrow p_r = \frac{\emptyset}{K}\mu \int_0^h v_r(z)\,dz = \frac{\emptyset}{K}\mu v_{r,0} h\left(\frac{\lambda+1}{\lambda+2}\right) \tag{E26}$$

Thus including the varying evaporation rate then the final differential equation becomes,

$$\frac{2\sigma \cos\theta}{R_c} = \rho_l g h + \frac{\emptyset}{K}\mu h \frac{dh}{dt} + \frac{\emptyset}{K}\mu\left(\frac{2\bar{\dot{m}}_e(W+t_p)}{\rho_l W t_p \emptyset}\right)\left(\frac{\lambda+1}{\lambda+2}\right)h^2 \tag{E27}$$

Multiply with $\frac{K}{\emptyset\mu h}$ on both sides, Eq. E27 becomes,

$$\frac{2\sigma \cos\theta}{\emptyset\mu h}\ \frac{K}{R_c} = \frac{\rho_l g K}{\emptyset\mu} + \frac{dh}{dt} + \left(\frac{2\bar{\dot{m}}_e(W+t_p)}{\rho_l W t_p \emptyset}\right)\left(\frac{\lambda+1}{\lambda+2}\right)h \tag{E28}$$

This can be represented as,

$$\frac{a}{h} = b + \dot{h} + ch \tag{E29}$$

Where the coefficients of a, b, and c ($\bar{\dot{m}}_e$ is substituted from Eq. 41) are defined as

$$a = \left(\frac{2\sigma\cos\theta}{\emptyset\mu}\right)\frac{K}{R_c}, \quad b = \frac{\rho_l g K}{\emptyset\mu}, \quad c = \frac{2*\left[\frac{k_L(T_{L-V}-\bar{T}_S)}{\frac{L_C}{2}} + \bar{h}(T_\infty - \bar{T}_S) + \sigma_S\epsilon\left(T_\infty^4 - \bar{T}_S^4\right)\right]*(W+t_p)(\lambda+1)}{h_{fg}\rho_l W t_p \emptyset(\lambda+2)}$$

The analytical answer for the time required to reach a specific height of the liquid front, denoted as $t(h)$, can be obtained by considering the effects of evaporation and gravity. The equation can be rewritten as follows.

$$\int \frac{h}{-ch^2 - bh + a}\,dh = \int 1 dt \tag{E30}$$

With the boundary condition of $z = 0$ at $t = 0$, the solution to Eq. E30 is given in the form of $t(h)$ rather than $h(t)$ (see Eq. E31) where,

The solution to the initial integral is provided through the utilization of the subsequent formulation. Defining $\beta = -4ac - b^2$ and for $\beta < 0$, the solution in terms of t = t (h) is,

$$t = \frac{1}{2c}\left[-\ln\left(\frac{-ch^2 - bh + a}{a}\right)\right] - \left[\frac{b}{2c\sqrt{-\beta}}\ln\left\{\frac{(-2ch - b - \sqrt{-\beta})(-b + \sqrt{-\beta})}{(-2ch - b + \sqrt{-\beta})(-b - \sqrt{-\beta})}\right\}\right] \tag{E31}$$

***-λ model***

From Figure 12, We can write differential mass balance as,

$$\dot{M}(z) = \dot{M}(z + dz - dz) \tag{E32}$$

$$\Rightarrow \dot{M}(z) = \dot{M}(z + dz) - d\dot{M}(z) \tag{E33}$$

$$\Rightarrow d\dot{M}(z) = -2\dot{m}_e\left(W + t_p\right)dz \tag{E34}$$

Here, the evaporation rate $\dot{m}_e$ is a function of z, then we can write it as,

$$\dot{m}_e = a_1\left(1 - \frac{z}{h}\right)^{-\lambda} \tag{E35}$$

To find evaporation constant '$a_1$' we integrate from 0 to h. we get

$$\int_0^h d\dot{M}(e) = \int_0^h -2\dot{m}_e\left(W + t_p\right)dz \quad \text{(E36)}$$

Total evaporative mass flow, $\dot{M}(e) = \int_0^h -2\dot{m}_e\left(W + t_p\right)dz$ (E37)

$$\Rightarrow \dot{M}(e) = \int_0^h -2a_1\left(1-\frac{z}{h}\right)^{-\lambda}\left(W + t_p\right)dz \quad \text{(E38)}$$

$$\Rightarrow -2\bar{\dot{m}}_e\left(W + t_p\right)h = -2a_1\frac{(W+t_p)}{(1-\lambda)}h \quad \text{(E39)}$$

$$\Rightarrow \bar{\dot{m}}_e = \frac{a_1}{(1-\lambda)} \quad \text{(E40)}$$

We substitute $a_1$ in Eq. E35 to get $\dot{m}_e = \bar{\dot{m}}_e(1-\lambda)\left(1-\frac{z}{h}\right)^{-\lambda}$ (E41)

When integrating and using the boundary condition that the total mass inflow at $z = 0$ must be equal to $\dot{M}_{\dot{h}} + \dot{M}_{\dot{e}}$, one obtains

$$\int d\dot{M}(z) = \int -2\dot{m}_e\left(W + t_p\right)dz \quad \text{(E42)}$$

$$\Rightarrow \dot{M}(z) = \int -2\bar{\dot{m}}_e(1-\lambda)\left(1-\frac{z}{h}\right)^{-\lambda}\left(W + t_p\right)dz \quad \text{(E43)}$$

$$\Rightarrow \dot{M}(z) = -2\bar{\dot{m}}_e(1-\lambda)\frac{(W+t_p)}{(1-\lambda)}(h-z)\left(1-\frac{z}{h}\right)^{-\lambda} + c \quad \text{(E44)}$$

$$\Rightarrow \dot{M}(z) = 2\bar{\dot{m}}_e\left(W + t_p\right)(h-z)\left(1-\frac{z}{h}\right)^{-\lambda} + c \quad \text{(E45)}$$

Applying the boundary condition that at $z = 0$; $\dot{M}(z) = \dot{M}_{\dot{h}} + \dot{M}(e)$

$$\dot{M}_{\dot{h}} + \dot{M}(e) = 2\bar{\dot{m}}_e\left(W + t_p\right)(h-0)(1-0)^{-\lambda} + c \quad \text{(E46)}$$

$$\dot{M}_{\dot{h}} + \dot{M}(e) = 2\bar{\dot{m}}_e\left(W + t_p\right)h + c \quad \text{(E47)}$$

$$\Rightarrow c = \dot{M}_{\dot{h}} + \dot{M}(e) - 2\bar{\dot{m}}_e\left(W + t_p\right)h \quad \text{(E48)}$$

Substituting ‘c’ in Eq. E45 we get,

$$\dot{M}(z) = 2\bar{\dot{m}}_e\left(W + t_p\right)(h-z)\left(1-\frac{z}{h}\right)^{-\lambda} + \dot{M}_{\dot{h}} + \dot{M}(e) - 2\bar{\dot{m}}_e\left(W + t_p\right)h \quad \text{(E49)}$$

We know that $\dot{M}(e) = 2\bar{\dot{m}}_e\left(W + t_p\right)h$

$$\dot{M}(z) = 2\bar{\dot{m}}_e\left(W + t_p\right)(h-z)\left(1-\frac{z}{h}\right)^{-\lambda} + \dot{M}_{\dot{h}} \quad \text{(E50)}$$

$$\Rightarrow \dot{M}(z) = \dot{M}_{\dot{h}} + 2\bar{\dot{m}}_e\left(W + t_p\right)h\left[1-\frac{z}{h}\right]^{1-\lambda} \quad \text{(E51)}$$

The local mass flow $\dot{M}(z)$ can be expressed as a function of flow velocity, which is subsequently utilized to determine the viscous pressure loss.

At the point where z equals zero, the velocity of the refill reaches its maximum value.

$$v_{r,0} = \frac{\dot{M}_e}{\rho_l A_b} = \frac{2\bar{\dot{m}}_e h(w+t_p)}{\rho_l w t_p \emptyset} \quad \text{(E52)}$$

Evaporation rate varies with height and the refill velocity becomes,

$$v_r(z) = v_{r,0}\left[1 - \frac{z}{h}\right]^{1-\lambda} \tag{E53}$$

The momentum balance of the liquid inside the FP is

$$p_c = p_h + p_{\dot{h}} + p_r \tag{E54}$$

Viscous pressure loss can be calculated as

$$p_{\dot{h}} = \frac{\emptyset}{K}\mu\int_0^h \dot{h}\, dz = \frac{\emptyset}{K}\mu h\dot{h} \tag{E55}$$

$$\Rightarrow p_r = \frac{\emptyset}{K}\mu\int_0^h v_r(z)\, dz = \frac{\emptyset}{K}\mu v_{r,0}\frac{h}{2-\lambda} \tag{E56}$$

Thus including the varying evaporation rate then the final differential equation becomes

$$\frac{2\sigma\cos\theta}{R_c} = \rho_l gh + \frac{\emptyset}{K}\mu h\frac{dh}{dt} + \frac{\emptyset}{K}\mu\left(\frac{2\bar{\dot{m}}_e(W+t_p)}{\rho_l W t_p \emptyset}\right)\frac{h^2}{2-\lambda} \tag{E57}$$

Multiply with $\frac{K}{\emptyset\mu h}$ on both sides of Eq. E57 becomes,

$$\frac{2\sigma\cos\theta}{\emptyset\mu h}\ \frac{K}{R_c} = \frac{\rho_l gK}{\emptyset\mu} + \frac{dh}{dt} + \left(\frac{2\bar{\dot{m}}_e(W+t_p)}{\rho_l W t_p \emptyset}\right)\frac{h}{2-\lambda} \tag{E58}$$

This can be represented as

$$\frac{a}{h} = b + \dot{h} + ch \tag{E59}$$

Where the coefficients of a, b, and c ($\bar{\dot{m}}_e$ is substituted from Eq. 41) are defined as

$$a = \left(\frac{2\sigma\cos\theta}{\emptyset\mu}\right)\frac{K}{R_c}\ ,\ b = \frac{\rho_l gK}{\emptyset\mu},\qquad c = \frac{2*\left[\frac{k_L(T_{L-V}-\bar{T}_S)}{\frac{L_C}{2}} + \bar{h}(T_\infty-\bar{T}_S) + \sigma_S\epsilon\left(T_\infty^4-\bar{T}_S^4\right)\right]*(W+t_p)}{h_{fg}\rho_l W t_p\emptyset(2-\lambda)}$$

The analytical answer for the time required to reach a specific height of the liquid front, denoted as $t(h)$, can be obtained by considering the effects of evaporation and gravity. The equation can be rewritten as follows.

$$\int\frac{h}{-ch^2-bh+a}dh = \int 1dt \tag{E60}$$

With the boundary condition of z = 0 at t = 0, the solution to Eq. E60 is given in the form of $t(h)$ rather than $h(t)$ (see Eq.E61) where,

The solution to the initial integral is provided through the utilization of the subsequent formulation. Defining $\beta = -4ac - b^2$ and for $\beta < 0$ the total solution in terms of t = t (h) is

$$t = \frac{1}{2c}\left[-\ln\left(\frac{-ch^2-bh+a}{a}\right)\right] - \left[\frac{b}{2c\sqrt{-\beta}}\ \ln\left\{\frac{(-2ch-b-\sqrt{-\beta})(-b+\sqrt{-\beta})}{(-2ch-b+\sqrt{-\beta})(-b-\sqrt{-\beta})}\right\}\right] \tag{E61}$$

**Appendix-B: Uncertainty analysis of all the measured and derived values**

The method of uncertainty analysis evaluates a derived quantity by considering the uncertainties present in the experimentally measured variables. This process identifies and assesses potential errors and combines the results to determine the overall uncertainty of the measurement. In this study, the uncertainties are computed using the fractional change approximation method. Key parameters influencing the results include evaporation rate, convective heat transfer coefficient, wicking length, ambient temperature, permeability, and liquid-vapor surface temperature. Table 5 presents the errors associated with the measurement of each individual parameter.

Table 5 Individual uncertainities in quantities for Whatman 1001FP, 1004FP, and 1005FP.

| **Sl. no.** | **Quantity** | **Uncertainty** | | |
|---|---|---|---|---|
| | | 1004 FP | 1001 FP | 1005 FP |
| 1 | Height | $0.08 mm$ | $0.09 mm$ | $0.09 mm$ |
| 2 | Permeability | $0.13 \times 10^{-13} m^2$ | $0.03 \times 10^{-13} m^2$ | $0.05 \times 10^{-14} m^2$ |
| 3 | Mass gain | $2.51 \times 10^{-7}$ kg s$^{-1}$ | – | – |
| 4 | Evaporative Mass loss | $9.70 \times 10^{-9}$ kg s$^{-1}$ | – | – |
| 5 | Cumulative mass loss | $3.86 \times 10^{-7}$ kg s$^{-1}$ | – | – |
| 6 | $T_{LV}$ | $0.05\ K$ | $0.05\ K$ | $0.05\ K$ |
| 7 | $T_\infty$ | $0.05\ K$ | $0.05\ K$ | $0.05K$ |
| 8 | $\bar{h}$ | $1.03\ W\ m^{-2}\ K^{-1}$ | $1.24\ W\ m^{-2}\ K^{-1}$ | $1.26\ W\ m^{-2}\ K^{-1}$ |
| 9 | $\bar{\dot{m}}_e$ | $1.29 \times 10^{-5} kg\ m^{-2}\ s^{-1}$ | $1.55 \times 10^{-5} kg\ m^{-2}\ s^{-1}$ | $1.52 \times 10^{-5} kg\ m^{-2}\ s^{-1}$ |

**Pixelation error in optical analysis**

i. $(\boldsymbol{h\ vs\ t})$

This research examines how the capillary penetration length (h) of liquid ethanol varies over time in three types of filter papers: 1004FP, 1001FP, and 1005FP. The measurements were recorded with a pixelation error of 0.08 mm/pixel for 1004FP, and 0.09 mm/pixel for both 1001FP and 1005FP.

i. $\boldsymbol{h^2\ vs\ t}$

Estimation of permeabilities involve plotting $h^2$ versus initial times (see Section 4) where the slope of linear $h^2 - t$ curve yields its value. It is important to get uncertainty in $h^2$ as well for this purpose as follows (and plotted in Figure 9),

$$U_h = d(h^2) = 2\ h\ dh = 2\ h\ U_h$$

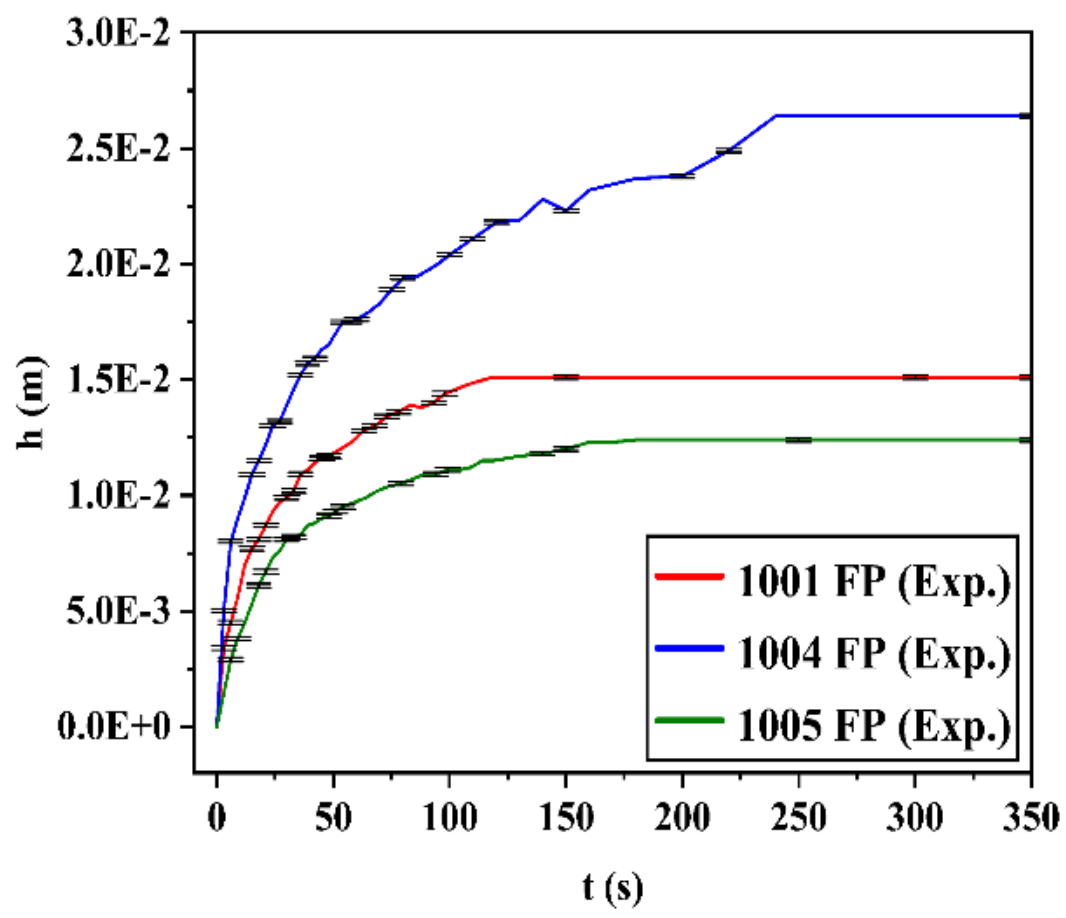

Figure 19 Temporal variation of capillary penetration length for liquid ethanol in three different filter papers. The error bar represents the pixelation error of 0.08mm/pixel for 1004FP and 0.09mm/pixel for 1001FP and 1005FP cases.

**Uncertainty analysis of + $\lambda$ and – $\lambda$ models:**

The uncertainty [26] in the calculated result '$t$' (see Eq. 36) was obtained using a direct computer-executed uncertainty analysis, involving sequential perturbation of input variables. The procedure followed is outlined below:

1. Calculate the initial result '$t$' using the recorded data, and store this value as $t_0$.
2. For each variable $X_i$ (where $i$ ranges from 1 to $N$, the total number of variables in $t$ in this case the number of variables are 5 namely $K, T_{LV}, \bar{h}$, $T_\infty$, and $\overline{T_s}$):
   i. Increase the value of $X_i$ by its uncertainty interval $\Delta X_i$ and calculate the resulting $\boldsymbol{t_{i+}}$ using this augmented value while keeping all other variables at their nominal values.
   ii. Compute the difference $t_{i+} - t_0$ and store this as $M_{i+}$ the contribution to the uncertainty in $t$ due to a positive perturbation of $X$.
   iii. As '$t$' is a nonlinear function of $Xi$ , decrease $X_i$ by $\Delta X_i$ and calculate the resulting $\boldsymbol{t_{i-}}$. Then, compute $M_{i-} = t_i - t_0$ .
   iv. Use the average of the absolute values of $M_{i+}$ and $M_{i-}$ as the final contribution $M_i$.
3. Calculate the total uncertainty in t as the root-sum-square (RSS) of the individual contributions $M_i$ as,

$$\text{Total Uncertainty} = \sqrt{\sum_{i=1}^{N}(M_i)^2} \qquad (44)$$

Thus, the total uncertainty for Eq. (36) is quantified for +λ and – λ models, with a maximum total uncertainty to be ~15% for 1005 case, and the maximum total uncertainties for all the FP cases are given in Table 6.

Table 6: Maximum total uncertainty due to uncertainty propagation for different filter papers using +λ and −λ models.

| Total uncertainty (%) | | |
|---|---|---|
| FP | +λ model | – λ model |
| 1001 FP | 11.37 | 9.09 |
| 1004 FP | 3.80 | 5.78 |
| 1005 FP | 14.52 | 15.21 |